%% file: main.tex
\documentclass[10pt,twocolumn,letterpaper]{extarticle}
\usepackage[T1]{fontenc}
\usepackage[tt=false,type1=true]{libertine}
\usepackage[varqu]{zi4}
\usepackage{amsmath,amsthm}
\usepackage[libertine]{newtxmath}
\usepackage{microtype}
\usepackage[letterpaper,left=54pt,right=54pt,top=58pt,bottom=62pt,columnsep=24pt]{geometry}
\usepackage{graphicx}
\usepackage{booktabs}
\usepackage{caption}
\usepackage[numbers,sort&compress]{natbib}
\usepackage{hyperref}
\hypersetup{hidelinks,
  pdftitle={Truthful and Expected-Payment-Optimal Model Routing in Graph-Based Agentic Workflows},
  pdfauthor={Jiayuan Liu, Mingyu Guo, Jiarui Gan, Vincent Conitzer},
  pdfkeywords={Mechanism design, procurement auctions, strategy-proofness, LLM routing, graph-based agentic workflows, context reuse}}
\usepackage{titlesec}
\titleformat{\section}{\fontsize{12}{14}\selectfont\bfseries\raggedright\hyphenpenalty=10000}{\thesection}{0.6em}{}
\titleformat{\subsection}{\fontsize{11}{13}\selectfont\bfseries\raggedright\hyphenpenalty=10000}{\thesubsection}{0.6em}{}
\titleformat{\paragraph}[runin]{\normalfont\itshape}{}{0pt}{}
\titlespacing*{\section}{0pt}{10pt plus 2pt minus 2pt}{4pt}
\titlespacing*{\subsection}{0pt}{8pt plus 2pt minus 2pt}{3pt}
\titlespacing*{\paragraph}{0pt}{5pt plus 1pt minus 1pt}{0.5em}
\newtheoremstyle{paperplain}{0.5\baselineskip}{0.5\baselineskip}
  {\itshape}{\parindent}{\scshape}{.}{0.5em}{}
\theoremstyle{paperplain}
\newtheorem{theorem}{Theorem}
\newtheorem{proposition}[theorem]{Proposition}
\newtheorem{corollary}[theorem]{Corollary}
\newcommand{\Description}[1]{}
\usepackage{algorithm}
\usepackage{algorithmic}
\usepackage{multirow}
\usepackage{nicefrac}
\usepackage{tikz}
\usetikzlibrary{positioning,arrows.meta,calc,shapes.geometric,backgrounds,fit}
\usepackage{enumitem}

\title{\bfseries\fontsize{17}{20}\selectfont
Truthful and Expected-Payment-Optimal Model Routing in Graph-Based Agentic Workflows}
\author{%
  \normalsize\textbf{Jiayuan Liu}\textsuperscript{1,2}\quad
  \textbf{Mingyu Guo}\textsuperscript{3}\quad
  \textbf{Jiarui Gan}\textsuperscript{4}\quad
  \textbf{Vincent Conitzer}\textsuperscript{1,2}\\[5pt]
  \small\textsuperscript{1}Carnegie Mellon University\quad
  \textsuperscript{2}Foundations of Cooperative AI Lab (FOCAL)\\[2pt]
  \small\textsuperscript{3}Adelaide University\quad
  \textsuperscript{4}University of Oxford\\[4pt]
  \small\texttt{jiayuan4@cs.cmu.edu}\quad
  \texttt{mingyu.guo@adelaide.edu.au}\quad
  \texttt{jiarui.gan@cs.ox.ac.uk}\quad
  \texttt{conitzer@cs.cmu.edu}}
\date{}
\begin{document}
\maketitle
\begin{abstract}
Assigning routine tasks to cheaper large language model (LLM) providers can reduce generation cost, but switching providers may require processing accumulated context again. Keeping dependent tasks with the same provider can instead reuse eligible context. Graph-based workflows therefore couple routing decisions across tasks. Independent providers may also privately know their real-time unit costs and strategically misreport them.
We formulate this routing problem as a single-parameter procurement auction and propose \textsc{Star} (\textbf{S}trategy-proof \textbf{T}opology-aware \textbf{A}gent \textbf{R}outing), which minimizes virtual-cost-weighted graph workloads. Its allocation rule and payments, computed by integrating workload over possible cost reports, make truthful reporting utility-maximizing regardless of other providers' reports and cover each truthful provider's execution cost. With independent cost distributions and non-decreasing virtual costs, \textsc{Star} additionally minimizes the user's expected total payment to providers. Controlled synthetic experiments separate the execution-cost benefit of topology awareness from the payment benefit of virtual-cost optimization: the former grows with fan-out and transferred payload, while the latter increases gradually along the tested prior-shape heterogeneity path. A paired sensitivity study quantifies how payment savings vary with prior misspecification while strategy-proofness is preserved.
\end{abstract}
\par\medskip

\begin{figure*}[t]
\centering
\includegraphics[width=\textwidth]{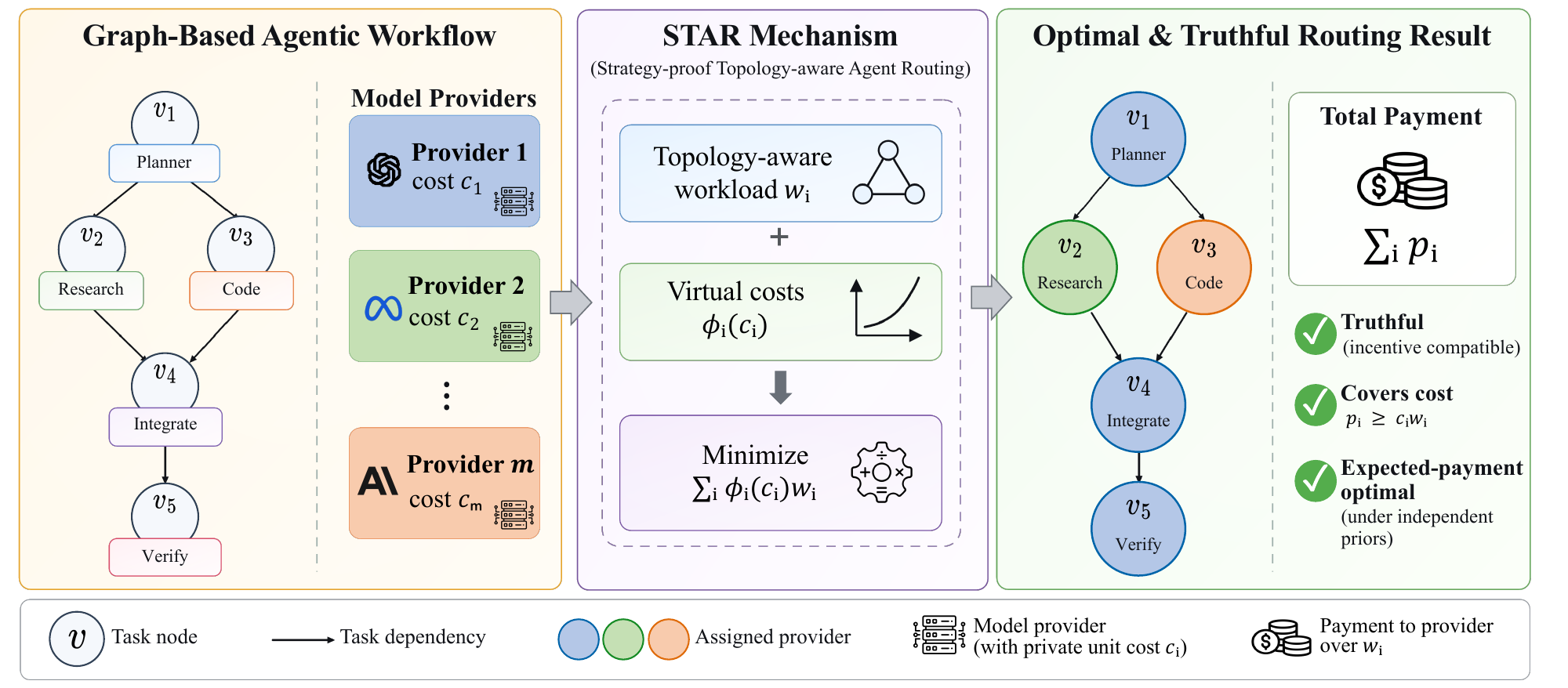}
\caption{\textbf{\textsc{Star} overview.} \textsc{Star} assigns a fixed workflow's tasks to eligible model providers and determines their payments. Providers privately know unit costs $c_i$; assignment-dependent workloads include generation and upstream-context processing with eligible reuse. \textsc{Star} uses reported costs $b_i$ to optimize virtual-cost-weighted workloads and compute payments that make truthful reporting utility-maximizing and cover truthful providers' execution costs. The figure uses $w_i=q_i(A)$ and, because of strategy-proofness, depicts truthful cost reports $b_i=c_i$. With independent, correctly specified cost priors and non-decreasing virtual costs, \textsc{Star} minimizes the user's expected total payment among truthful, cost-covering mechanisms.}
\Description{Overview of a graph-based workflow with planner, research, code, integration, and verification tasks; independent providers with private unit costs; \textsc{Star}'s topology-aware, virtual-cost-weighted optimization; and the resulting task assignment and provider payments. Node colors identify assigned providers.}
\label{fig:star_overview}
\end{figure*}

\section{Introduction}

Large language model (LLM) systems increasingly organize model calls into dependent workflows. Frameworks such as Graph of Thoughts \citep{besta2024graph}, MetaGPT \citep{hong2023metagpt}, and ChatDev \citep{qian2024chatdev} use structured interactions to address complex tasks. Routing can reserve more capable models for demanding tasks and assign routine tasks to cheaper models that meet their requirements, reducing the cost of executing the workflow \citep{chen2024frugalgpt,ong2024routellm,zhang2025let}.

We consider a user who needs a workflow executed and selects a model provider for each task node. Providers independently supply model services and seek to maximize profit. Before they submit bids, the orchestrator identifies the providers that meet each task's requirements and specifies how their inference work is measured. Provider $i$ privately knows its unit cost $c_i$: the cost of one unit of this measured work. Multiplying $c_i$ by its assigned workload gives its execution cost. The user seeks to minimize the total payment to providers while satisfying the workflow's service requirements. Two features shape this routing problem.

First, a task's cost depends on the providers assigned to neighboring tasks. Its provider must generate output and process the context received from upstream tasks. Keeping adjacent tasks with the same provider can reuse cached input; switching providers requires fresh processing. For example, a coding assistant may use a more capable, more expensive model for planning, switch to a cheaper, less capable model for a routine edit it can handle, and return to the more capable model for verification. With separate serving contexts and no eligible cache reuse for the transferred context, the cheaper model processes the accumulated context, and the more capable model processes the updated context on return. These two context-processing charges can exceed the saving on the routine edit. Globally economical routing therefore balances cheaper task execution against the cost of context switches, especially when a large context feeds many downstream tasks.

Second, providers may privately know their instantaneous unit execution costs. These costs can vary with hardware utilization, available capacity, electricity prices, and local demand.\footnote{Capacity-sensitive compute pricing is already illustrated by AWS EC2 Spot Instances (\url{https://aws.amazon.com/ec2/spot/}) and Google Cloud Spot VMs (\url{https://cloud.google.com/solutions/spot-vms}), which advertise discounts of up to 90\% and 91\%, respectively, relative to regular on-demand prices.} If paid their submitted bids, providers have incentives to quote costs above their actual costs. A truthful mechanism jointly designs task assignments and payments so that reporting the actual cost maximizes each provider's utility.

We propose \textsc{Star}, a strategy-proof mechanism that assigns workflow tasks and pays providers using their reported unit costs and public graph-dependent workloads. It scores providers by the procurement analogue of Myerson's virtual value: the \emph{virtual cost} \(\phi_i(c)=c+F_i(c)/f_i(c)\), where \(F_i\) and \(f_i\) are the cumulative distribution function (CDF) and density of provider \(i\)'s cost~\citep{myerson1981optimal}. Its cost prior is the probability distribution of its private unit cost. The added term accounts for the payment incentive associated with that distribution. Under normalized truthful payments, expected total payment equals expected virtual-cost-weighted workload. Minimizing this objective for each reported cost profile therefore minimizes the user's expected total payment under independent regular priors, meaning independent cost distributions with non-decreasing virtual costs.

\textsc{Star} is \emph{dominant-strategy incentive compatible} (DSIC): reporting the true unit cost maximizes a provider's utility for any reports by the others. This strategy-proofness follows from a simple allocation property: increasing a provider's reported unit cost can only decrease or preserve its assigned workload. Comparing the optimal assignments before and after the increase proves this property even when neighboring assignments and cache reuse change together. Integrating the allocated workload over possible cost reports yields truthful payments and \emph{ex-post individual rationality} (IR): each truthful provider receives at least its execution cost~\citep{archer2001truthful}. Virtual-cost scoring selects the expected-payment-optimal allocation within this truthful framework.

Figure~\ref{fig:star_overview} summarizes this direct-report mechanism: providers submit their unit costs, and \textsc{Star} jointly computes task assignments and payments.

We distinguish two sources of economic improvement. Topology awareness can reduce realized execution cost by avoiding repeated fresh-input processing. Separately, virtual-cost optimization can reduce the user's expected payment relative to a graph-aware raw-score truthful mechanism. This second effect reflects differences in providers' cost distributions: provider-specific priors induce different virtual-cost markups and can change the globally preferred workflow allocation.

\paragraph{Contributions.}
First, we model task assignment and provider payments through graph-dependent workloads that capture the trade-off between cheaper models and repeated context processing. Second, we establish strategy-proofness and expected-payment optimality using the single-parameter framework, give exact allocation formulations, and derive an efficient payment formula for affine scores. Third, our controlled experiments quantify execution savings from topology awareness, payment savings from heterogeneous priors, and their sensitivity to prior specification, together with the runtime of the complete mechanism.

\section{Related Work}
\label{sec:related_work}
We provide a condensed overview here. Appendix~\ref{app:detailed_related_work} gives a comprehensive discussion.

\textbf{LLM orchestration and routing.}
Chain-, tree-, and graph-structured prompting and multi-agent frameworks organize LLM calls into dependent workflows~\citep{wei2022chain,yao2023tree,besta2024graph,wu2024autogen,hong2023metagpt,qian2024chatdev}. Cost-aware routers and cascades select among heterogeneous models~\citep{chen2024frugalgpt,ong2024routellm,zhang2025let}. MasRouter jointly selects collaboration modes, roles, and LLMs~\citep{yue2025masrouter}; BAMAS selects model pools and collaboration topologies under a budget~\citep{yang2026bamas}. Our setting adds strategic provider cost reports and truthful payments to routing on a public workflow.

\textbf{Mechanism design.}
Our mechanism uses procurement virtual costs from Myerson's optimal-auction framework~\citep{myerson1981optimal} and the single-parameter monotonicity/payment characterization of~\citet{archer2001truthful}. Prior work studies truthful cloud-resource and crowdsourcing procurement~\citep{zaman2013combinatorial,singla2013truthful}, and frugal procurement over public set systems including paths and vertex covers~\citep{chen2010frugal}. Our model accounts for workload changes across the complete task assignment and optimizes the user's expected payment. Public workload functions connect these graph interactions to the single-parameter incentive framework.

\textbf{Context reuse.}
PagedAttention, SGLang, and prompt caching reduce repeated context processing~\citep{kwon2023efficient,zheng2023efficiently,gim2024prompt}. We represent the applicable reuse discount by a public edge-level cached-input fraction \(\kappa_{u,v}\), determined by cache eligibility and the serving configuration.

\section{System Model}
\label{sec:model}

\subsection{Workflow and Feasible Allocations}

We represent a workflow as a finite directed acyclic graph (DAG) $\mathcal{G}=(\mathcal{V},\mathcal{E})$. Let $v_{\mathrm{user}}\in\mathcal{V}$ be a virtual source node representing the initial user input, and let
$\mathcal{V}_{T}
=
\mathcal{V}\setminus\{v_{\mathrm{user}}\}$
be the set of task nodes.
There are $N$ providers $\mathcal{M}=\{1,\ldots,N\}$. For every task node $v$, the public set $\mathcal{M}_v\subseteq\mathcal{M}$ specifies the providers that are capable of executing $v$. An allocation is a mapping
$A:\mathcal{V}_{T}\rightarrow\mathcal{M}$
such that $A(v)\in\mathcal{M}_v$ for every $v\in\mathcal{V}_{T}$. Let $\mathcal{A}$ denote the bid-independent set of all feasible allocations. For notational convenience, we set $A(v_{\mathrm{user}})=0$, where $0\notin\mathcal{M}$. We partition the dependency edges into source and internal edges:
\begin{equation}
\mathcal E_{\mathrm{src}}
=
\{(v_{\mathrm{user}},v)\in\mathcal E\},
\qquad
\mathcal E_{\mathrm{int}}
=
\mathcal E\setminus\mathcal E_{\mathrm{src}}.
\label{eq:edge_partition}
\end{equation}

All token quantities below are measured in millions of tokens. Let $L_v^{\mathrm{out}}$ denote the expected generated output at task node $v$, and let $L_{u,v}^{\mathrm{trans}}$ denote the expected payload transmitted along edge $(u,v)$. In particular, $L_{v_{\mathrm{user}},v}^{\mathrm{trans}}$ is the initial user input provided to a root task.

\subsection{Public Graph-Dependent Workloads}

For allocation $A$, define
$
x_{v,i}(A)
=
\mathbf{1}\{A(v)=i\}$.
We set $x_{v_{\mathrm{user}},i}(A)=0$. For every edge $(u,v)$, define
\begin{equation}
\delta_{u,v,i}(A)
=
x_{v,i}(A)\bigl(1-x_{u,i}(A)\bigr).
\label{eq:cache_miss_indicator}
\end{equation}
Thus, $\delta_{u,v,i}(A)=1$ exactly when provider $i$ executes $v$ but does not execute its parent $u$.

Provider $i$ has a public input-to-output workload coefficient $\rho_i$. For each edge $(u,v)$, let $\kappa_{u,v}\in[0,1]$ denote its public cached-input fraction: $\kappa_{u,v}=0$ represents free cached input, whereas $\kappa_{u,v}=1$ represents no same-provider discount. We set $\kappa_{u,v}=1$ for edges that are not cache eligible.

\paragraph{Running example.}
Consider a hub $h$ and two children fixed to provider $H$. Each outgoing edge carries payload $L$, with discount $\kappa$. Assigning $h$ to $H$ contributes $2\rho_H\kappa L$ to the children's input workload; assigning it to another provider contributes $2\rho_H L$. The saved input work, $2\rho_H(1-\kappa)L$, must be weighed against the hub's output generation and incoming-input processing. Proposition~\ref{prop:broadcaster_threshold} accounts for both in the corresponding score threshold.

The total standardized workload assigned to provider $i$ under allocation $A$ is
\begin{equation}
\begin{aligned}
q_i&(A)
={}
\sum_{v\in\mathcal{V}_{T}}
L_v^{\mathrm{out}}x_{v,i}(A)
\\
&+
\rho_i
\sum_{(u,v)\in\mathcal{E}}
L_{u,v}^{\mathrm{trans}}
\left[
\kappa_{u,v}x_{v,i}(A)
+
(1-\kappa_{u,v})\delta_{u,v,i}(A)
\right].
\end{aligned}
\label{eq:provider_workload}
\end{equation}

For an internal edge whose endpoints are both assigned to provider $i$, the edge contributes $\rho_i\kappa_{u,v}L_{u,v}^{\mathrm{trans}}$. If the provider changes, the same edge contributes the full fresh-input workload $\rho_iL_{u,v}^{\mathrm{trans}}$. Since the virtual source is not assigned to any provider, every source-to-root payload is charged as fresh input.

Equation~\eqref{eq:provider_workload} measures input-processing work using public cache eligibility and discounts: $\kappa_{u,v}=1$ charges full fresh-input work, while smaller values represent eligible reuse.

\subsection{Private Costs, Reports, and Utilities}

Provider $i$ privately knows its unit cost, a single number
$c_i\in[a_i,d_i]$
where $[a_i,d_i]$ is its publicly known range of possible costs. This number is the cost per unit of the publicly measured workload $q_i(A)$, so its realized execution cost under allocation $A$ is
$C_i(A;c_i)=c_iq_i(A)$.
Provider $i$ submits a report $b_i\in[a_i,d_i]$. Given payment $p_i(\mathbf b)$, its quasi-linear utility is
\begin{equation}
u_i(\mathbf b;c_i)
=
p_i(\mathbf b)
-
c_iq_i(A(\mathbf b)).
\label{eq:provider_utility}
\end{equation}

The provider's private information is the single number $c_i$. Its complete workload function $q_i:\mathcal{A}\rightarrow\mathbb{R}_{\geq0}$, token quantities, $\rho_i$, cache eligibility, $\kappa_{u,v}$, and feasible-provider sets $\mathcal{M}_v$ are public and fixed before reports are submitted. These public service specifications define the single-parameter model used throughout the incentive analysis.

\section{\textsc{Star}: Strategy-Proof Topology-Aware Routing}
\label{sec:method}

\subsection{Score-Based Graph Allocation}

We first define a general score-based allocation rule. For each provider $i$, let $s_i:[a_i,d_i]\rightarrow\mathbb{R}$ be a public non-decreasing score function. Given reports $\mathbf b=(b_1,\ldots,b_N)$, define
\begin{equation}
A^{s}(\mathbf b)
\in
\arg\min_{A\in\mathcal{A}}
\sum_{j\in\mathcal{M}}
s_j(b_j)q_j(A).
\label{eq:score_allocation}
\end{equation}
When multiple allocations attain the same primary objective, the mechanism uses a fixed deterministic tie-breaking order that is independent of all reports.

For \textsc{Star}, provider costs are independently drawn from provider-specific distributions with CDF $F_i$ and density $f_i$ on $[a_i,d_i]$. The procurement virtual cost is defined as
\begin{equation}
\phi_i(c)
=
c+\frac{F_i(c)}{f_i(c)}.
\label{eq:virtual_cost}
\end{equation}
We assume regularity, so that $\phi_i$ is non-decreasing. \textsc{Star} instantiates Eq.~\eqref{eq:score_allocation} with $s_i=\phi_i$:
\begin{equation}
A^{\textsc{Star}}(\mathbf b)
\in
\arg\min_{A\in\mathcal{A}}
\sum_i
\phi_i(b_i)q_i(A).
\label{eq:star_allocation}
\end{equation}

As a special case, for the uniform prior $c_i\sim\operatorname{Uniform}[a_i,d_i]$ used in our reference configuration,
$\phi_i(c)=2c-a_i$.

The term $F_i(c)/f_i(c)$ adjusts the allocation score for the provider's cost distribution. The payment rule determines its realized information rent, the payment in excess of execution cost: $p_i-c_iq_i(A)$.

\subsection{Workload Monotonicity}

We prove workload monotonicity by comparing the optimal assignments at two reported costs. The graph determines the workload functions and thus the objective minimized in each comparison, incorporating all neighboring assignments and cache effects.

\begin{proposition}[Workload Monotonicity]
\label{prop:graph_monotonicity}
Fix provider $i$ and the other reports $\mathbf b_{-i}$. Suppose that $s_i$ is non-decreasing and that Eq.~\eqref{eq:score_allocation} is solved exactly using the fixed report-independent tie-breaking rule. Then
\begin{equation}
q_i\!\left(A^s(b_i,\mathbf b_{-i})\right)
\end{equation}
is non-increasing in $b_i$.
\end{proposition}

\paragraph{Proof.}
Take two reports $z<z'$ and let
\[
A=A^s(z,\mathbf b_{-i}),
\qquad
A'=A^s(z',\mathbf b_{-i}).
\]
Define the contribution of all other providers as
\begin{equation}
R_i(B)
=
\sum_{j\neq i}s_j(b_j)q_j(B).
\end{equation}
Optimality of $A$ at report $z$ gives
\begin{equation}
s_i(z)q_i(A)+R_i(A)
\leq
s_i(z)q_i(A')+R_i(A').
\label{eq:mono_first}
\end{equation}
Optimality of $A'$ at report $z'$ gives
\begin{equation}
s_i(z')q_i(A')+R_i(A')
\leq
s_i(z')q_i(A)+R_i(A).
\label{eq:mono_second}
\end{equation}
Adding Eqs.~\eqref{eq:mono_first} and~\eqref{eq:mono_second} yields
\begin{equation}
\bigl(s_i(z')-s_i(z)\bigr)
\bigl(q_i(A')-q_i(A)\bigr)
\leq 0.
\end{equation}
If $s_i(z')>s_i(z)$, then $q_i(A')\leq q_i(A)$. If the two scores are equal, the two optimization problems have the same objective, and the fixed report-independent tie-breaking rule selects the same allocation. Hence the selected workload is non-increasing in the report. \hfill$\square$

The proof covers changes to several neighboring assignments at once. Their effects on cache reuse and provider workloads are captured by the functions $q_j(A)$ and the contribution $R_i(A)$ of the other providers.

\subsection{Payments, Strategy-Proofness, and Individual Rationality}

For fixed $\mathbf b_{-i}$, abbreviate
$q_i(z)=q_i\!\left(A^s(z,\mathbf b_{-i})\right)$.
For any public non-decreasing score functions, we compute a provider's payment by adding its reported execution cost to the area under its workload curve over higher possible cost reports:
\begin{equation}
p_i(\mathbf b)
=
b_iq_i(b_i)
+
\int_{b_i}^{d_i}q_i(z)\,dz.
\label{eq:integral_payment}
\end{equation}

Equation~\eqref{eq:integral_payment} pays each marginal unit of workload at its critical cost report: the highest reported cost at which that unit remains assigned to the provider. The integral is an area under a workload curve; payments are monetary amounts.

\begin{theorem}[DSIC and Ex-Post Individual Rationality]
\label{thm:dsic_ir}
Under the conditions of Proposition~\ref{prop:graph_monotonicity}, the score-based allocation rule in Eq.~\eqref{eq:score_allocation} together with the payment rule in Eq.~\eqref{eq:integral_payment} is dominant-strategy incentive compatible and ex-post individually rational.
\end{theorem}

\paragraph{Proof.}
Fix provider $i$, its true cost $c_i$, and the other reports $\mathbf b_{-i}$. Its truthful utility is
\begin{equation}
u_i(c_i,\mathbf b_{-i};c_i)
=
\int_{c_i}^{d_i}q_i(z)\,dz
\geq 0,
\end{equation}
which establishes ex-post individual rationality.

For an alternative report $r_i>c_i$,
\begin{align}
&u_i(c_i,\mathbf b_{-i};c_i)
-
u_i(r_i,\mathbf b_{-i};c_i)
\notag\\
&\qquad=
\int_{c_i}^{r_i}
\bigl(q_i(z)-q_i(r_i)\bigr)\,dz
\geq 0,
\end{align}
because $q_i$ is non-increasing. For $r_i<c_i$,
\begin{align}
&u_i(c_i,\mathbf b_{-i};c_i)
-
u_i(r_i,\mathbf b_{-i};c_i)
\notag\\
&\qquad=
\int_{r_i}^{c_i}
\bigl(q_i(r_i)-q_i(z)\bigr)\,dz
\geq 0.
\end{align}
Thus truthful reporting maximizes provider $i$'s utility for every fixed $\mathbf b_{-i}$. \hfill$\square$

Theorem~\ref{thm:dsic_ir} is the standard single-parameter payment characterization applied to the graph-dependent workload in Eq.~\eqref{eq:provider_workload} \citep{archer2001truthful}.

\subsection{Expected-Payment Optimality}
\label{sec:expected_payment_optimality}

The preceding truthfulness result applies to any non-decreasing score function. The particular choice $s_i=\phi_i$ provides the additional expected-payment guarantee.

\begin{theorem}[Expected-Payment Optimality]
\label{thm:expected_payment}
Suppose costs are independent, provider $i$ has support $[a_i,d_i]$, every virtual cost $\phi_i$ is non-decreasing, and payments use the normalization $u_i(d_i,\mathbf c_{-i};d_i)=0$. Then \textsc{Star} minimizes
$\mathbb{E}_{\mathbf c}\left[\sum_i p_i(\mathbf c)\right]$
among deterministic DSIC and ex-post individually rational mechanisms whose allocations lie in $\mathcal{A}$.
\end{theorem}

\paragraph{Proof.}
Fix provider $i$ and the other costs $\mathbf c_{-i}$, and write $q_i(c)=q_i(A(c,\mathbf c_{-i}))$. Independence leaves $i$'s conditional cost distribution equal to $F_i$. Taking the expectation of the payment integral and exchanging the order of integration gives
\begin{align}
\mathbb E_{c_i}\!\left[\int_{c_i}^{d_i}q_i(z)\,dz\right]
&=\int_{a_i}^{d_i}q_i(z)\!\left(\int_{a_i}^{z}f_i(c)\,dc\right)dz\notag\\
&=\int_{a_i}^{d_i}F_i(z)q_i(z)\,dz.
\label{eq:expected_rent_identity}
\end{align}
Thus the expected payment, conditional on $\mathbf c_{-i}$, is
\begin{align}
\mathbb E_{c_i}[p_i(c_i,\mathbf c_{-i})]
&=\int_{a_i}^{d_i}\bigl[cf_i(c)+F_i(c)\bigr]q_i(c)\,dc\notag\\
&=\mathbb E_{c_i}[\phi_i(c_i)q_i(c_i)].
\label{eq:conditional_virtual_payment}
\end{align}
Taking expectation over the other costs and summing over providers yields
\[\mathbb{E}_{\mathbf c}
\Big[
\sum_i p_i(\mathbf c)
\Big]
=
\mathbb{E}_{\mathbf c}
\Big[
\sum_i\phi_i(c_i)q_i(A(\mathbf c))
\Big].\]

Equation~\eqref{eq:star_allocation} minimizes the expression inside the expectation pointwise for every type profile. Regularity makes the virtual scores non-decreasing, so Proposition~\ref{prop:graph_monotonicity} and Theorem~\ref{thm:dsic_ir} make this minimizing allocation truthful and individually rational. For any DSIC allocation, the payment characterization adds the provider's utility at its upper cost endpoint $d_i$ to the normalized payment. Ex-post individual rationality makes this boundary utility nonnegative, so setting it to zero minimizes payments for the same allocation. Together these facts establish expected-payment optimality. Appendix~\ref{app:expected_payment_identity} records the normalization details. \hfill$\square$

\paragraph{Execution cost, virtual objective, and payment.}
We distinguish three economic quantities:
\begin{align}
C_{\mathrm{exec}}(A;\mathbf c)
&=
\sum_i c_iq_i(A),
\label{eq:execution_cost}
\\
J_{\phi}(A;\mathbf b)
&=
\sum_i\phi_i(b_i)q_i(A),
\label{eq:virtual_objective}
\\
P(\mathbf b)
&=
\sum_i p_i(\mathbf b).
\label{eq:total_payment}
\end{align}
\textsc{Star} minimizes $J_\phi$ for each reported profile and hence minimizes the user's expected total payment $\mathbb{E}[P]$ under the correct priors. Execution cost $C_{\mathrm{exec}}$ measures providers' resource expenditure, whereas $P$ measures the money the user pays them; the payment includes compensation for execution and incentive rent.

\begin{corollary}[Raw-Score DSIC]
\label{cor:raw_dsic}
Setting $s_i(b_i)=b_i$ in Eq.~\eqref{eq:score_allocation} and using Eq.~\eqref{eq:integral_payment} yields a DSIC and ex-post individually rational mechanism. Under truthful reports, its allocation minimizes realized execution cost:
\begin{equation}
A^{\mathrm{raw}}(\mathbf c)
\in
\arg\min_{A\in\mathcal A}
\sum_i c_iq_i(A).
\end{equation}
\end{corollary}

\begin{corollary}[Misspecified Monotone Scores]
\label{cor:misspecified_scores}
Any public non-decreasing score functions preserve DSIC and ex-post IR with the payment rule in Eq.~\eqref{eq:integral_payment}. Using the true virtual costs additionally gives expected-payment optimality under Theorem~\ref{thm:expected_payment}.
\end{corollary}

\subsection{Exact Allocation}
\label{sec:exact_allocation}

We give a mixed-integer linear program (MILP) for finite directed acyclic graphs (DAGs), used in all reported experiments, followed by an exact dynamic program for rooted trees and chains. Appendix~\ref{app:tree_decomp} extends dynamic programming to graphs of bounded treewidth~\citep{cygan2015parameterized}.

Using the edge partition in Eq.~\eqref{eq:edge_partition}, let $x_{v,i}\in\{0,1\}$ indicate that task node $v\in\mathcal V_T$ is assigned to provider $i\in\mathcal M$. For every edge $(u,v)\in\mathcal E$, let $\delta_{u,v,i}\in[0,1]$ indicate that $v$ is assigned to provider $i$ while its parent $u$ is not assigned to $i$.

For generic provider scores $s_i(b_i)$, the allocation problem is
\begin{align}
\min_{x,\delta}\quad
&\sum_{i\in\mathcal M}s_i(b_i)\Bigg[\sum_{v\in\mathcal V_T}L_v^{\mathrm{out}}x_{v,i}\notag\\
&\qquad+\rho_i\sum_{(u,v)\in\mathcal E}L_{u,v}^{\mathrm{trans}}\Bigl(\kappa_{u,v}x_{v,i}+(1-\kappa_{u,v})\delta_{u,v,i}\Bigr)\Bigg]
\label{eq:milp_objective}\\
\text{s.t.}\quad
&\sum_{i\in\mathcal M_v}x_{v,i}=1,\quad v\in\mathcal V_T,
\label{eq:milp_assignment}\\
&x_{v,i}=0,\quad v\in\mathcal V_T,\ i\notin\mathcal M_v,
\label{eq:milp_capability}\\
&\delta_{u,v,i}\geq x_{v,i}-x_{u,i},\quad (u,v)\in\mathcal E_{\mathrm{int}},\ i\in\mathcal M,
\label{eq:milp_delta_lower}\\
&\delta_{u,v,i}\leq x_{v,i},\quad (u,v)\in\mathcal E_{\mathrm{int}},\ i\in\mathcal M,
\label{eq:milp_delta_upper_child}\\
&\delta_{u,v,i}\leq1-x_{u,i},\quad (u,v)\in\mathcal E_{\mathrm{int}},\ i\in\mathcal M,
\label{eq:milp_delta_upper_parent}\\
&\delta_{v_{\mathrm{user}},v,i}=x_{v,i},\quad (v_{\mathrm{user}},v)\in\mathcal E_{\mathrm{src}},\ i\in\mathcal M,
\label{eq:milp_source_edge}\\
&x_{v,i}\in\{0,1\},\quad v\in\mathcal V_T,\ i\in\mathcal M,\notag\\
&0\leq\delta_{u,v,i}\leq1,\quad (u,v)\in\mathcal E,\ i\in\mathcal M.
\label{eq:milp_domains}
\end{align}
The objective is the score-weighted workload in Eq.~\eqref{eq:provider_workload}; Eqs.~\eqref{eq:milp_assignment}--\eqref{eq:milp_capability} assign each task to exactly one eligible provider. Constraints~\eqref{eq:milp_delta_lower}--\eqref{eq:milp_delta_upper_parent} enforce $\delta_{u,v,i}=\allowbreak x_{v,i}(1-x_{u,i})$. Thus a shared provider processes the cached-input fraction $\kappa_{u,v}$, while a provider switch charges full fresh-input workload. Equation~\eqref{eq:milp_source_edge} charges every source-to-root payload as fresh input.

\textsc{Star} uses $s_i(b_i)=\phi_i(b_i)$ and Raw-DSIC uses $s_i(b_i)=b_i$. Both optimize the same feasible assignments and graph-dependent workloads, differing only in their provider scores.

Equal-objective allocations are resolved by the lexicographic order of public node ordering and provider identifiers. This tie order is fixed before reports are submitted.

All reported allocations are computed with the COIN-OR Branch-and-Cut (CBC) optimization solver through the Python modeling library PuLP, requesting zero relative and absolute optimality gap. We retain solutions with solver status \texttt{Optimal}, using documented floating-point tolerances. Exact score minimization supplies the workload monotonicity required by Proposition~\ref{prop:graph_monotonicity}; an approximation-based extension would likewise need a monotone allocation rule to preserve strategy-proofness.

\paragraph{Exact optimization on rooted trees.}
For rooted trees and chains, exact dynamic programming combines optimal child assignments from leaves to root, accounting for their connecting edges and fresh user input. It computes the optimal objective in $O(|\mathcal V_T|N^2)$ operations; Appendix~\ref{app:tree_dp} gives the recurrence, correctness proof, assignment reconstruction, and extension to forests.

\subsection{A Computational Special Case: Affine Scores}
\label{sec:affine_payment_special_case}

For affine scores, which are linear functions of the reported cost plus a constant, \(s_i(b)=\alpha_i b+\gamma_i\) with \(\alpha_i>0\), define
\begin{equation}
J_i(t;\mathbf b_{-i})=\min_{A\in\mathcal A}\left[tq_i(A)+\sum_{j\ne i}s_j(b_j)q_j(A)\right].
\label{eq:parametric_value}
\end{equation}
The payment in Eq.~\eqref{eq:integral_payment} then equals
\begin{equation}
p_i(\mathbf b)=b_iq_i(A^s(\mathbf b))+\frac{J_i(s_i(d_i);\mathbf b_{-i})-J_i(s_i(b_i);\mathbf b_{-i})}{\alpha_i}.
\label{eq:affine_two_solve_payment}
\end{equation}
This formula evaluates the payment integral through repeated optimization of the same task-assignment problem. We first find the optimal assignment at the submitted costs; its objective value is reused for every provider's payment. For each provider assigned positive workload, we then recompute the optimum with that provider's report set to its upper endpoint $d_i$, holding the other reports fixed. Thus all payments need one shared optimization and at most one additional optimization per active provider, plus tie-breaking computations. Raw, uniform, and power-prior scores have slopes \(1\), \(2\), and \(1+1/k_i\), respectively. Appendix~\ref{app:affine_payment_derivation} proves the optimal-value identity and specifies these score families.

\subsection{When Topology Awareness Matters}

\begin{proposition}[Broadcaster Threshold]
\label{prop:broadcaster_threshold}
Consider a hub node $h$ with $m$ outgoing edges of equal payload $L^{\mathrm{trans}}$ and cached-input fraction $\kappa$ to children assigned to provider $H$. Hold every other task's assignment fixed, and compare assigning $h$ to eligible providers $L$ and $H$, with scores $s_L$ and $s_H$. Let $I_L$ and $I_H$ be the score-weighted costs of processing all input arriving at $h$ under these two choices, including their respective cache discounts, and write $\Delta_{\mathrm{in}}=I_H-I_L$. Assigning $h$ to $H$ is strictly preferred exactly when
\begin{equation}
m(1-\kappa)\rho_HL^{\mathrm{trans}}s_H
>
L_h^{\mathrm{out}}(s_H-s_L)+\Delta_{\mathrm{in}}.
\label{eq:broadcaster_threshold}
\end{equation}
\end{proposition}

\paragraph{Proof.}
Assigning $h$ to $L$ contributes
$s_LL_h^{\mathrm{out}}
+
m\rho_HL^{\mathrm{trans}}s_H+I_L$,
whereas assigning it to $H$ contributes
$
s_HL_h^{\mathrm{out}}
+
m\kappa\rho_HL^{\mathrm{trans}}s_H+I_H$.
The latter is smaller exactly under Eq.~\eqref{eq:broadcaster_threshold}. \hfill$\square$

The left side is the saved input-processing cost at the children; the right side is the extra generation and incoming-input cost at the hub. When $h$ has no incoming input, $\Delta_{\mathrm{in}}=0$. At equality, the fixed tie-breaking rule applies. The outgoing-context saving grows linearly with fan-out and payload and is zero at $\kappa=1$. Both \textsc{Star} and graph-aware raw-score routing capture this trade-off; virtual-cost scoring further accounts for provider-specific payment incentives.

\section{Experiments}
\label{sec:experiments}

We validate the incentive and payment implementation, separate execution-cost gains from payment gains, and test prior-shape heterogeneity, deliberate prior misspecification, and payload/cache controls. Appendix~\ref{app:runtime_details} reports computational costs.

\subsection{Experimental Setup}

\paragraph{Structural-prior stress suite.}
Our primary payment study uses a frozen broadcaster-market construction and three pre-specified regular-prior profiles. Each provider follows the power family
\begin{equation}
F_i(c)=\left(\frac{c-a_i}{d_i-a_i}\right)^{k_i},\qquad c\in[a_i,d_i],
\label{eq:stress_power_prior}
\end{equation}
whose virtual cost is given in Appendix~\ref{app:affine_payment_derivation}. The shape parameter \(k_i>0\) controls how probability is distributed within the provider's cost range: \(k_i=1\) is uniform, \(k_i>1\) places more probability near the upper endpoint, and \(k_i<1\) places more near the lower endpoint. The mild and strong profiles use fixed provider-specific \(k_i\) vectors. For each profile, we evaluate the same 24 pre-specified structural cells with 50 paired holdout trials per cell. Provider cost ranges, shape vectors, capability tiers, and the structural grid are specified in Appendix~\ref{app:stress_config}.

\paragraph{Scope and workflows.}
Four hand-constructed synthetic DAG families inspired by creative writing, software engineering, web development, and legal analysis provide the S1--S4 reference. They vary fan-out, fan-in, and high-bandwidth broadcaster motifs, with eligibility specified by manually assigned capability tiers. Appendix~\ref{app:market_config} gives each workflow's graph and task requirements.

\paragraph{Synthetic workflow reference.}
S1--S4 use nine providers in three capability tiers, with fixed base-cost anchors in a $1{:}5{:}25$ progression and uniform cost draws within provider-specific ranges. We set $\rho_i=1/3$, $\kappa_{u,v}=0.1$ on cache-eligible edges, and $\kappa_{u,v}=1$ otherwise. Nominal token quantities receive lognormal variation with $\sigma=0.3$; the S2 source context follows a truncated-Pareto distribution on $5\mathrm{k}$--$100\mathrm{k}$ tokens. Appendix~\ref{app:market_config} gives the cost and token distributions and base-cost anchors.

\paragraph{Mechanisms and routing diagnostics.}
\textsc{Star} minimizes the full graph-dependent virtual-cost objective and uses Eq.~\eqref{eq:integral_payment}, evaluated by Eq.~\eqref{eq:affine_two_solve_payment} for affine scores. Raw-DSIC minimizes \(\sum_i b_iq_i(A)\) over the same allocation space and applies the corresponding payment rule; under truthful reports, its allocation minimizes realized execution cost and coincides with the true-cost oracle. The True-Cost Oracle directly minimizes \(\sum_i c_iq_i(A)\), serving as an execution-cost benchmark. Myopic independently selects the cheapest feasible provider at each node. We compare \textsc{Star} and Raw-DSIC by payments to providers and execution costs. The true-cost oracle provides the execution-cost benchmark, while Myopic is evaluated by execution cost only.

\paragraph{Metrics.}
We separately report realized execution cost \(C_{\mathrm{exec}}=\sum_i c_iq_i(A)\), procurement payment \(P=\sum_i p_i\), and information rent \(R=P-C_{\mathrm{exec}}\). If \(C_{\mathrm{oracle}}\) denotes the true-cost oracle objective, then
\begin{equation}
P_{\textsc{Star}}-C_{\mathrm{oracle}}=\underbrace{C_{\textsc{Star}}-C_{\mathrm{oracle}}}_{\text{allocation distortion}}+\underbrace{P_{\textsc{Star}}-C_{\textsc{Star}}}_{\text{information rent}}.
\label{eq:payment_decomposition}
\end{equation}
We use the same random draws for each mechanism in a comparison and report paired-bootstrap 95\% confidence intervals (CIs).

\begin{figure*}[t]
\centering
\includegraphics[width=0.9\textwidth]{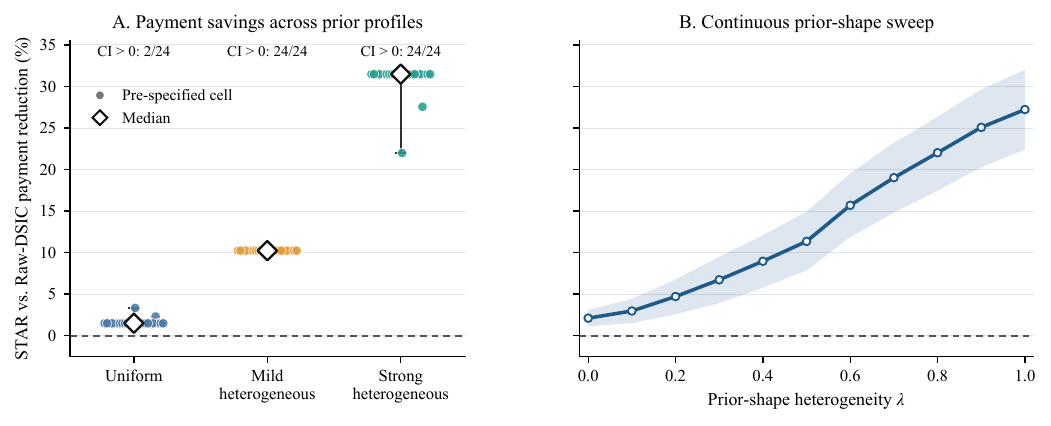}
\caption{\textbf{Payment savings under heterogeneous regular priors.} \textbf{A:} Each point is one of the 24 pre-specified structural cells for the indicated prior profile; the larger marker and range summarize the cell-level point estimates, and the annotation reports how many paired 95\% intervals are strictly above zero. \textbf{B:} In the representative degree-100 high-payload graph, \textsc{Star}'s payment reduction along the continuous prior-shape path. Provider supports, graph semantics, and latent cost quantiles remain fixed; the band denotes a paired-bootstrap 95\% confidence interval.}
\Description{Two-panel plot. Panel A compares \textsc{Star} payment reductions across uniform, mildly heterogeneous, and strongly heterogeneous prior profiles over 24 structural cells. Panel B plots payment reduction with a confidence band as prior-shape heterogeneity increases from zero to one.}
\label{fig:heterogeneous_prior_main}
\end{figure*}

\subsection{Implementation Validation}

The theorems establish DSIC and ex-post individual rationality; exhaustive small-instance enumeration and dense bid sweeps validate the implementation. The final suite contains 159 passing tests. Across all reported exact experiment rows, CBC returns status \texttt{Optimal} with zero reported gap; when providers report their actual costs, Raw-DSIC assigns each task to exactly the provider selected by the execution-cost oracle; the minimum truthful provider utility is zero; and all payment-decomposition checks pass up to floating-point tolerance. The monotonicity case study evaluates 901 bids and contains four workload plateaus. Workload never increases with the bid, and no sampled misreport gives higher utility than reporting the actual cost.

\subsection{Payment Savings under Heterogeneous Regular Priors}
\label{sec:heterogeneous_priors}

We first isolate the condition that distinguishes \textsc{Star} from Raw-DSIC: provider-specific regular cost distributions. Both mechanisms optimize the same graph-dependent workload and use the DSIC payment rule, with Raw-DSIC scoring provider $i$ by $b_i$ and \textsc{Star} scoring it by $\phi_i(b_i)$. We evaluate three pre-specified prior profiles over the same 24 frozen structural cells, using 50 paired holdout trials per cell and retaining every cell regardless of effect size.

\input{tables/heterogeneous_prior_summary.tex}

Table~\ref{tab:heterogeneous_prior_summary} shows consistent payment savings across all 24 structural cells for the tested mild and strong prior-shape profiles. Under the strong profile, \textsc{Star} reduces payment relative to Raw-DSIC by \(21.99\%\)--\(31.48\%\), and all 24 paired 95\% confidence intervals lie strictly above zero. The mild profile yields reductions of \(10.23\%\)--\(10.30\%\), again with positive intervals in all 24 cells. The uniform profile yields \(1.49\%\)--\(3.34\%\), with a strictly positive interval in 2 of 24 cells.

In the representative degree-100 high-payload cell with \(\rho=1/3\), the payment reductions are 1.49\% with 95\% CI: $[-0.39\%,\,3.58\%]$ under the uniform profile, 10.23\%, $[3.44\%,16.73\%]$ under the mild profile, and 31.48\%, $[22.56\%,40.73\%]$ under the strong profile. The fixed synthetic distributions enable a paired comparison of prior shape while preserving the same graph and workload rules.

\input{tables/e2_misspecification_main.tex}

\subsection{Continuous Prior-Shape Heterogeneity}
\label{sec:prior_continuum}

To test whether the heterogeneous-prior gain depends on a selected endpoint, we hold each provider's support \([a_i,d_i]\) fixed and vary only its regular prior shape. For \(\lambda\in\{0,0.1,\ldots,1\}\), define
\begin{equation}
\begin{aligned}
k_i(\lambda)&=\left(k_i^{\mathrm{strong}}\right)^\lambda,\\
F_i(c;\lambda)&=\left(\frac{c-a_i}{d_i-a_i}\right)^{k_i(\lambda)}.
\end{aligned}
\label{eq:prior_continuum}
\end{equation}
The corresponding virtual cost is
\begin{equation}
\phi_i(c;\lambda)=\left(1+\frac{1}{k_i(\lambda)}\right)c-\frac{a_i}{k_i(\lambda)}.
\label{eq:prior_continuum_virtual}
\end{equation}
We use 200 paired holdout trials per \(\lambda\) and common latent cost quantiles throughout the sweep.

Figure~\ref{fig:heterogeneous_prior_main}B shows a gradual increase in payment savings across the prior-shape path. The payment reduction rises from \(2.11\%\) at \(\lambda=0\) to \(27.22\%\) at \(\lambda=1\). The point estimate increases at all ten adjacent grid steps and forms an approximately smooth curve, with the final step adding \(2.14\) percentage points.

For each cost draw, a rank reversal occurs when a pair of providers eligible for the same task have opposite orderings under raw and virtual costs. Counting each eligible pair once, the fraction with a reversal increases from \(0.072\) to \(0.171\) along the path. Across the eleven grid points, payment reduction has Spearman correlation \(0.998\) with this rate and \(0.936\) with the fraction of draws in which \textsc{Star} and Raw-DSIC choose different task assignments. These descriptive associations summarize how allocation decisions and payment savings co-vary along the tested synthetic path.

At \(\lambda=0\), all providers have uniform-shaped priors with different lower support endpoints \(a_i\). The virtual scores \(\phi_i(c)=2c-a_i\) therefore contain provider-specific offsets, which can change the preferred allocation and produce a nonzero \textsc{Star}/Raw-DSIC difference at this endpoint.

At \(\lambda=1\), mean \((P,C_{\mathrm{exec}},R)\) is \((1.8382,1.4530,0.3851)\) for \textsc{Star} and \((2.5256,1.2587,1.2670)\) for Raw-DSIC; the oracle execution cost is \(1.2587\). Hence the absolute payment difference is \(0.6875\): \textsc{Star} reduces information rent by \(0.8818\) while increasing execution cost by \(0.1944\). The payment saving is primarily lower incentive rent, partially offset by allocation distortion.

The degree-10, degree-20, and degree-100 continuum instances replicate the same consumer structure. Their coincident normalized payment-reduction and unique-pair rank-reversal curves provide a consistent view of the prior-shape sweep at three scales. Increasing degree scales the absolute saving from \(0.276\) at degree 10 to \(0.687\) at degree 100 when \(\lambda=1\).

\subsection{Sensitivity to Prior-Estimation Error Magnitude}
\label{sec:robustness_runtime}

We test how \textsc{Star}'s payment advantage changes as errors in estimated provider cost distributions increase. True costs follow the power family in Eq.~\eqref{eq:stress_power_prior}: the uniform profile has $k_i=1$ for every provider, while the heterogeneous power-prior profile has provider-specific shapes $k=(0.5,2,0.2,5,1)$ (Appendix~\ref{app:stress_config}). We change only the assumed shapes, $\widehat k_i=k_i e^{\delta r_i}$, with fixed $r=(1,-1,0,1,-1)$ and both signs of $|\delta|\in\{0.05,0.1,0.2\}$. These magnitudes permit shape increases of at most $5.13\%$, $10.52\%$, and $22.14\%$, respectively. We pre-specify all six perturbations and 1,000 new cost vectors per true profile. Each condition shares the same true costs, truthful reports, graph, eligibility, and cost ranges with correct-prior \textsc{Star} and Raw-DSIC.

Table~\ref{tab:e2_misspecification_main} reports correct-prior savings of $0.79\%$ (uniform) and $30.04\%$ (heterogeneous). Under heterogeneous power priors, all six perturbations retain payment reductions of $29.92\%$--$30.16\%$, with every paired 95\% interval strictly above zero. Under uniform priors, all six sample-mean reductions remain positive: $0.72\%$--$0.85\%$ at $|\delta|=0.05$, $0.56\%$--$0.71\%$ at $|\delta|=0.1$, and $0.20\%$--$0.34\%$ at $|\delta|=0.2$. Both directions at $|\delta|=0.05$ and $0.1$ have intervals strictly above zero. At $|\delta|=0.2$, both intervals span zero, indicating less precise evidence for these smaller savings. The grid characterizes the payment benefit and its precision as prior-estimation error grows. All tested assumed scores are increasing and used consistently in allocation and payments, preserving strategy-proofness. Correctly specified priors additionally yield the expected-payment optimum in Theorem~\ref{thm:expected_payment}. At $|\delta|=0.2$, the entire task assignment matches correct-prior routing in at least $93\%$ of uniform draws and $96.9\%$ of heterogeneous draws in each error direction. Appendix~\ref{app:runtime_details} reports the full assignment-stability and payment-decomposition results.

\subsection{Topology Stress Tests and Structural Controls}
\label{sec:topology_controls}

\input{tables/structural_controls.tex}

Table~\ref{tab:structural_controls} isolates the roles of payload and reuse at fixed degree 100 and \(\rho=1/3\), matching Proposition~\ref{prop:broadcaster_threshold}. Raising the transferred-context/child-output ratio from one to sixteen increases execution-cost reduction from 2.13\% to 67.47\%; removing reuse with \(\kappa=1\) makes it exactly zero. The paired intervals identify repeated fresh-input processing as the source of the large topology gains.

Topology awareness and virtual-cost optimization generate distinct effects. In the tested degree-100 high-payload fan-out construction, graph-aware raw-cost routing reduces realized execution cost by \(90.20\%\)--\(96.91\%\) relative to Myopic routing, depending on \(\rho\) and the prior profile. Comparing \textsc{Star} with the already topology-aware Raw-DSIC then isolates the additional payment benefit of provider-specific virtual-cost scoring.

Appendices~\ref{app:uniform_reference}, \ref{app:stress_results}, \ref{app:prior_continuum}, and~\ref{app:support_width} report the synthetic S1--S4 uniform-prior reference, the complete stress grid, prior-continuum diagnostics, and the support-width comparison, respectively.

\section{Conclusion and Future Directions}
\label{sec:conclusion}

\textsc{Star} combines topology-aware workflow assignment with truthful provider payments. Workload monotonicity and the workload-integrating payment rule establish DSIC and ex-post IR; independent, correctly specified regular cost priors yield expected-payment optimality. Controlled synthetic experiments distinguish execution savings from avoided context processing and payment savings from virtual costs, with prior-shape sweeps and paired misspecification tests characterizing the latter.

Affine scores enable efficient payment computation. Learned cost priors, measured serving workloads, and richer private service attributes provide directions for future work.

\section*{Acknowledgments}
Jiayuan Liu and Vincent Conitzer thank the Cooperative AI Foundation, Macroscopic Ventures (formerly Polaris Ventures / the Center for Emerging Risk Research) and Jaan Tallinn's donor-advised fund at Founders Pledge for financial support.

\label{end:mainbody}
\clearpage
\label{start:references}
\bibliographystyle{abbrvnat}
\bibliography{references}

\clearpage

\appendix
\label{start:appendices}

\section{Detailed Related Work}
\label{app:detailed_related_work}

\paragraph{Structured LLM workflows and economical routing.}
Chain-, tree-, and graph-based reasoning and multi-agent systems structure dependent model calls~\citep{wei2022chain,yao2023tree,besta2024graph,wu2024autogen,hong2023metagpt,qian2024chatdev}. FrugalGPT, RouteLLM, and economical model routing select among models using performance and cost information~\citep{chen2024frugalgpt,ong2024routellm,zhang2025let}. Building on this motivation for economical model selection, our formulation incorporates independently compensated providers and incentives for truthful cost reporting.

\paragraph{Joint multi-agent configuration.}
MasRouter selects collaboration modes, roles, and LLMs in a cascaded router~\citep{yue2025masrouter}. BAMAS jointly selects a model pool and collaboration topology subject to a budget~\citep{yang2026bamas}. These systems demonstrate joint optimization of multi-agent configurations. Our complementary procurement setting fixes the workflow and eligibility sets before bidding, then uses strategic provider cost reports to determine task assignments and payments.

\paragraph{Single-parameter procurement.}
Our incentive analysis applies the single-parameter characterization for providers whose private information is one unit-cost value~\citep{archer2001truthful}. For any bid-independent feasible set and public quantities $q_i(A)$, exact minimization of $\sum_i s_i(b_i)q_i(A)$ with monotone scores and fixed tie-breaking yields workload monotonicity. Public quantity functions incorporate graph interactions directly into this argument. Procurement virtual costs then provide the expected-payment objective under independent regular priors~\citep{myerson1981optimal,nisan2007algorithmic}. We develop the graph-dependent workload formulation, exact allocation methods, payment computation, and economic comparisons within this framework.

\paragraph{Graph and infrastructure procurement.}
Truthful cloud-resource allocation and crowdsourcing procurement study related incentive problems~\citep{zaman2013combinatorial,singla2013truthful}. Frugal mechanism design on public set systems includes path and vertex-cover procurement~\citep{chen2010frugal}; its overpayment benchmarks and worst-case frugality objectives differ from our correct-prior expected-payment objective. A provider here can execute multiple tasks, and its public amount of work changes with neighboring assignments and input-reuse terms. Our public quantity functions represent this allocation-dependent work while preserving a single private unit cost per provider. The same framework also accommodates public setup- and placement-dependent workload rules.

\paragraph{Context processing and caching.}
PagedAttention, SGLang, and prompt caching motivate attention to context-processing work~\citep{kwon2023efficient,zheng2023efficiently,gim2024prompt}. Actual reuse depends on the model, serving instance, cache state, and token-prefix structure. Our workload model encodes the applicable eligibility and discount in public edge parameters, with $\kappa_{u,v}=1$ charging fresh input. The synthetic experiments fix these parameters to isolate the routing and payment effects of repeated context processing.

\section{Expected-Payment Identity}
\label{app:expected_payment_identity}

Section~\ref{sec:expected_payment_optimality} proves the expected-payment identity by exchanging the order of integration in the payment rule. This appendix records how the choice of upper-type utility enters that identity. For fixed other types $\mathbf c_{-i}$, write $q_i(z)=q_i(A(z,\mathbf c_{-i}))$ and let $U_i^{\mathrm{top}}(\mathbf c_{-i})$ be provider $i$'s utility when its true cost and report are both $d_i$. The single-parameter payment characterization gives
\begin{equation}
p_i(c_i,\mathbf c_{-i})=c_iq_i(c_i)+\int_{c_i}^{d_i}q_i(z)\,dz+U_i^{\mathrm{top}}(\mathbf c_{-i}).
\end{equation}
Ex-post individual rationality implies $U_i^{\mathrm{top}}(\mathbf c_{-i})\geq0$. The normalized rule sets it to zero and minimizes payment for the given allocation. With a general upper-type utility, the identity proved in the main text becomes $\mathbb E[p_i]=\mathbb E[\phi_i(c_i)q_i(A(\mathbf c))]+\mathbb E[U_i^{\mathrm{top}}(\mathbf c_{-i})]$. This separates the allocation-dependent virtual-cost term from the nonnegative normalization term.

\section{Derivation of the Affine-Score Payment Shortcut}
\label{app:affine_payment_derivation}

Equation~\eqref{eq:integral_payment} defines the general truthful payment for any monotone score-based allocation rule. We now give a computational specialization for the case in which provider $i$'s score is affine in its report:
$s_i(b)
=
\alpha_i b+\gamma_i,
\
\alpha_i>0$.
Affine scores permit the payment integral to be evaluated using optimal-value differences, while Theorem~\ref{thm:dsic_ir} establishes the incentive guarantee for general monotone scores.
For fixed $\mathbf b_{-i}$, define
\begin{equation}
J_i(t;\mathbf b_{-i})
=
\min_{A\in\mathcal A}
\big[
tq_i(A)
+
\sum_{j\neq i}s_j(b_j)q_j(A)
\big].
\label{eq:parametric_value_appendix}
\end{equation}
Then the general integral payment in Eq.~\eqref{eq:integral_payment} has the exact affine-score representation
\begin{equation}
p_i(\mathbf b)
=
b_iq_i\!\left(A^s(\mathbf b)\right)
+
\frac{
J_i(s_i(d_i);\mathbf b_{-i})
-
J_i(s_i(b_i);\mathbf b_{-i})
}{
\alpha_i
}.
\label{eq:affine_two_solve_payment_appendix}
\end{equation}

All score families used in our reported experiments are affine. In particular:
\begin{align}
\text{Raw score:}\
&s_i(b)=b,\ (\alpha_i,\gamma_i)=(1,0),
\label{eq:raw_affine_score}\\
\text{Uniform prior:}\
&\phi_i(b)=2b-a_i,\ (\alpha_i,\gamma_i)=(2,-a_i),
\label{eq:uniform_affine_score}\\
\text{Power prior:}\
&\phi_i(b)=\left(1+\frac{1}{k_i}\right)b-\frac{a_i}{k_i},
\notag\\
&\hspace{1.8em}(\alpha_i,\gamma_i)
=\left(1+\frac{1}{k_i},-\frac{a_i}{k_i}\right).
\label{eq:power_affine_score}
\end{align}

The allocation is solved once at the submitted profile. For each provider with positive allocated workload, we solve one additional allocation problem with that provider's score set to its upper-support value and apply Eq.~\eqref{eq:affine_two_solve_payment}. If the current workload is zero, monotonicity implies that all higher-report workloads are also zero, so the provider's normalized payment is zero.

For a general regular prior with a non-affine virtual cost, Eq.~\eqref{eq:integral_payment} directly supplies the truthful payment used in Theorems~\ref{thm:dsic_ir} and~\ref{thm:expected_payment}. Our implementation and reported experiments use the affine families above, enabling the optimal-value shortcut in Eq.~\eqref{eq:affine_two_solve_payment}.

The main paper defines the general normalized payment as
\begin{equation}
p_i(\mathbf b)
=
b_iq_i\!\left(A^s(\mathbf b)\right)
+
\int_{b_i}^{d_i}
q_i\!\left(A^s(z,\mathbf b_{-i})\right)\,dz.
\label{eq:supp_general_integral_payment}
\end{equation}
This section derives a computational shortcut for the special case
\begin{equation}
s_i(b)=\alpha_i b+\gamma_i,
\qquad
\alpha_i>0.
\end{equation}

For fixed $\mathbf b_{-i}$, define
\begin{equation}
J_i(t;\mathbf b_{-i})
=
\min_{A\in\mathcal A}
\left[
tq_i(A)
+
\sum_{j\neq i}s_j(b_j)q_j(A)
\right].
\label{eq:supp_parametric_value}
\end{equation}
As the pointwise minimum of affine functions of $t$, $J_i$ is concave, piecewise linear, and absolutely continuous. At almost every $t$, its derivative is the workload of an optimal allocation:
\begin{equation}
\frac{\partial J_i(t;\mathbf b_{-i})}{\partial t}
=
q_i\!\left(A_i(t;\mathbf b_{-i})\right),
\label{eq:supp_value_derivative}
\end{equation}
where
\[
A_i(t;\mathbf b_{-i})
\in
\arg\min_{A\in\mathcal A}
\left[
tq_i(A)
+
\sum_{j\neq i}s_j(b_j)q_j(A)
\right].
\]
The fixed report-independent tie rule determines the selected optimizer at breakpoints; these finitely many points do not affect the integral.

Therefore,
\begin{align}
&J_i(s_i(d_i);\mathbf b_{-i})
-
J_i(s_i(b_i);\mathbf b_{-i})
\notag\\
&\qquad=
\int_{s_i(b_i)}^{s_i(d_i)}
q_i\!\left(A_i(t;\mathbf b_{-i})\right)\,dt
\notag\\
&\qquad=
\alpha_i
\int_{b_i}^{d_i}
q_i\!\left(A^s(z,\mathbf b_{-i})\right)\,dz,
\label{eq:supp_affine_envelope}
\end{align}
where the final equality uses $t=s_i(z)=\alpha_i z+\gamma_i$ and $dt=\alpha_i\,dz$. Substituting Eq.~\eqref{eq:supp_affine_envelope} into Eq.~\eqref{eq:supp_general_integral_payment} yields
\begin{equation}
p_i(\mathbf b)
=
b_iq_i\!\left(A^s(\mathbf b)\right)
+
\frac{
J_i(s_i(d_i);\mathbf b_{-i})
-
J_i(s_i(b_i);\mathbf b_{-i})
}{
\alpha_i
}.
\end{equation}

This identity evaluates the general truthful payment exactly for affine scores. For arbitrary monotone scores, the workload integral in Eq.~\eqref{eq:supp_general_integral_payment} remains the applicable payment formula.

\section{Relation of Raw-Score DSIC to Vickrey--Clarke--Groves Payments}
\label{app:raw_vcg}

The raw-score mechanism selects
\begin{equation}
A^{\mathrm{raw}}(\mathbf b)
\in
\arg\min_{A\in\mathcal A}
\sum_i b_iq_i(A)
\end{equation}
and uses the normalized single-parameter integral payment. The Vickrey--Clarke--Groves (VCG) rule compensates a provider according to the reduction in the other providers' reported costs made possible by its participation.

When provider $i$ can be excluded and reporting its upper type induces zero workload,
\begin{equation}
q_i(A^{\mathrm{raw}}(d_i,\mathbf b_{-i}))=0,
\end{equation}
the payment can be written as
\begin{align}
p_i^{\mathrm{raw}}(\mathbf b)
={}&
\min_{A:q_i(A)=0}
\sum_{j\neq i}b_jq_j(A)
\notag\\
&-
\sum_{j\neq i}
b_jq_j(A^{\mathrm{raw}}(\mathbf b)),
\end{align}
which is the Clarke-pivot VCG payment for the efficient allocation.

A provider may also retain unavoidable workload at its upper type, for example when it is the sole feasible provider for a task. The top-type-normalized payment accommodates this case and preserves DSIC and ex-post individual rationality, using the upper-type assignment as its normalization reference.

\section{Equivalent Breakpoint Representation}
\label{app:breakpoint_representation}

We give an equivalent mathematical breakpoint representation of the general workload-integrating payment. The reported affine-score experiments compute the same payment through optimal-value differences.

For fixed $\mathbf b_{-i}$, the workload $q_i(z)$ is a non-increasing step function. When $b_i<d_i$, partition the integration range at every interior workload change:
\[
b_i=t_0<t_1<\cdots<t_K=d_i.
\]
Let $Q_k$ be the constant workload on the open interval $(t_{k-1},t_k)$. Then Eq.~\eqref{eq:integral_payment} becomes
\begin{equation}
p_i(\mathbf b)
=b_iq_i(b_i)+\sum_{k=1}^{K}Q_k(t_k-t_{k-1}).
\label{eq:interval_payment}
\end{equation}
The first term uses the workload actually selected by the fixed tie-breaking rule at the current report, including when $b_i$ is a change point. Values at the other individual change points and at $d_i$ do not affect the integral. When $b_i=d_i$, the integration range is empty and the payment is $d_iq_i(d_i)$. Workload changes can arise from task reassignment or cache-status changes; the number of intervals need not equal the number of tasks currently assigned to provider $i$.

\section{Exact Dynamic Programming for Rooted Trees}
\label{app:tree_dp}

Section~\ref{sec:exact_allocation} uses an exact dynamic program for rooted trees and chains. We give its value recurrence and correctness proof before detailing assignment reconstruction and workflow forests.

For each task node $v$ and eligible provider $i$, define $D_v(i)$ as the minimum contribution to the allocation objective from $v$ and all its descendant tasks, when $v$ is assigned to $i$. It includes their output generation and input processing on edges between those tasks. An edge from $v$'s parent is accounted for when combining $v$ with that parent; fresh user-context processing is included at the root $r$. Thus the complete workflow optimum is $\min_{i\in\mathcal M_r}D_r(i)$. We compute these values from leaves to root by dynamic programming.

Write $s_i=s_i(b_i)$ and let $\operatorname{ch}(v)$ denote $v$'s children. The local generation contribution is $g_v(i)=s_iL_v^{\mathrm{out}}$ for $v\ne r$; at the root, $g_r(i)=s_i[L_r^{\mathrm{out}}+\rho_iL_{v_{\mathrm{user}},r}^{\mathrm{trans}}]$ also includes fresh source input. The edge contribution $\eta_{v,u}(i,j)=\allowbreak s_j\rho_jL_{v,u}^{\mathrm{trans}}[\kappa_{v,u}+(1-\kappa_{v,u})\mathbf 1\{i\ne j\}]$ accounts for processing the parent--child context. The recurrence is
\begin{equation}
\begin{aligned}
D_v(i)&=g_v(i)\\
&+\sum_{u\in\operatorname{ch}(v)}\min_{j\in\mathcal M_u}
\{D_u(j)+\eta_{v,u}(i,j)\}.
\end{aligned}
\label{eq:tree_dp_recurrence}
\end{equation}

\paragraph{Correctness.}
With $v$'s provider fixed, its child branches can be optimized independently. Each minimizing term combines the child's best assignment with its connecting edge, charging that edge once. At a leaf, $D_v(i)=g_v(i)$; induction up the tree establishes every entry and hence the optimal root value. Computing all values takes $O(|\mathcal V_T|N^2)$ operations. Backtracking recovers an optimal assignment with the public tie order.

\paragraph{Root choice and reconstruction.}
Equivalently, the context-processing edge factor can be written as
\begin{equation}
\eta_{v,u}(i,j)=s_j(b_j)\rho_jL_{v,u}^{\mathrm{trans}}
\bigl[\kappa_{v,u}+(1-\kappa_{v,u})\mathbf 1\{i\ne j\}\bigr].
\label{eq:tree_edge_factor}
\end{equation}
The root entry already includes fresh user-context processing, so the complete objective is
\begin{equation}
\min_{i\in\mathcal M_r}D_r(i).
\label{eq:tree_dp_root}
\end{equation}
A workflow forest with several task roots is handled by applying Eq.~\eqref{eq:tree_dp_root} independently to each root and summing the resulting objectives.

For reconstruction, each dynamic-programming entry stores its minimizing child-provider choices. Backtracking from an optimal root choice determines the assignment of every task. Among equal-objective subtree assignments, the public node and provider ordering selects the lexicographically smallest assignment; this ordering is fixed before reports are submitted, just as in the main mechanism.

Evaluating the objective values requires
\begin{equation}
O\left(\sum_{v\in\mathcal V_T}|\mathcal M_v|+
\sum_{(v,u)\in\mathcal E_{\mathrm{int}}}|\mathcal M_v||\mathcal M_u|\right)
\end{equation}
operations, bounded by $O(|\mathcal V_T|N^2)$ for a task tree. The first term accounts for node values, including single-node trees; the second enumerates eligible parent--child provider pairs.

\section{Exact Tree-Decomposition Dynamic Programming}
\label{app:tree_decomp}

This section gives an exact junction-tree dynamic program for finite DAGs whose underlying undirected task graph has bounded treewidth. A directed DAG may induce cycles in this underlying undirected graph, for example through multi-parent fan-in. This is distinct from a directed control-flow cycle: workflows with directed loops must first be unrolled into finite DAGs.

\paragraph{Pairwise factorization.}
Let
\[
\mathcal H=(\mathcal V_T,\mathcal E_H)
\]
be the underlying undirected task graph, where $\{u,v\}\in\mathcal E_H$ whenever either $(u,v)$ or $(v,u)$ is an internal dependency edge. Each task $v$ has assignment domain $\mathcal M_v$.

The source-edge and generation contributions can be combined into the unary factor
\begin{equation}
\psi_v(i)
=
s_i(b_i)
\left[
L_v^{\mathrm{out}}
+
\rho_i
\sum_{(v_{\mathrm{user}},v)\in\mathcal E_{\mathrm{src}}}
L_{v_{\mathrm{user}},v}^{\mathrm{trans}}
\right],
\qquad
i\in\mathcal M_v.
\label{eq:jt_unary_factor}
\end{equation}
For every internal directed edge $(u,v)$, define the pairwise factor
\begin{equation}
\psi_{u\rightarrow v}(i,j)
=
s_j(b_j)\rho_jL_{u,v}^{\mathrm{trans}}
\left[
\kappa_{u,v}
+
(1-\kappa_{u,v})\mathbf 1\{i\neq j\}
\right],
\label{eq:jt_pairwise_factor}
\end{equation}
where $i\in\mathcal M_u$ and $j\in\mathcal M_v$. The complete score objective is therefore
\begin{equation}
\sum_{v\in\mathcal V_T}
\psi_v(A(v))
+
\sum_{(u,v)\in\mathcal E_{\mathrm{int}}}
\psi_{u\rightarrow v}(A(u),A(v)).
\label{eq:jt_factorized_objective}
\end{equation}
Equation~\eqref{eq:jt_factorized_objective} is exactly the objective in Eq.~\eqref{eq:score_allocation}, expressed as unary and pairwise factors.

\paragraph{Tree decomposition.}
A tree decomposition of $\mathcal H$ is a tree $\mathcal T=(\mathcal I,\mathcal F)$ together with bags $\{B_t\subseteq\mathcal V_T:t\in\mathcal I\}$ satisfying:

\begin{enumerate}
    \item for every vertex $v$, the bags containing $v$ form a connected
    subtree of $\mathcal T$;
    \item for every edge $\{u,v\}\in\mathcal E_H$, at least one bag contains
    both $u$ and $v$.
\end{enumerate}

The width is
\[
w=\max_{t\in\mathcal I}|B_t|-1.
\]
We root $\mathcal T$ at an arbitrary bag $t_0$. Each unary and pairwise factor in Eq.~\eqref{eq:jt_factorized_objective} is assigned to exactly one bag containing all variables in its scope. For example, one may assign each factor to the highest such bag under the rooted decomposition.

Let $\mathcal Q_t$ be the factors assigned to bag $B_t$. For an assignment
\[
\mathbf a_t
\in
\prod_{v\in B_t}\mathcal M_v,
\]
define the local bag cost
\begin{equation}
\Psi_t(\mathbf a_t)
=
\sum_{\psi\in\mathcal Q_t}
\psi\!\left(
\mathbf a_t\big|_{\operatorname{scope}(\psi)}
\right).
\end{equation}
Because every factor is assigned to exactly one bag, no subtraction or double-counting correction is required.

For every non-root bag $t$, let $p(t)$ be its parent and define the separator
\[
S_t=B_t\cap B_{p(t)}.
\]
The message from bag $t$ to its parent is
\begin{align}
m_{t\rightarrow p(t)}(\mathbf a_{S_t})
=
\min_{\mathbf a_{B_t\setminus S_t}}
\Bigg[
\Psi_t(\mathbf a_t)
+
\sum_{c\in\operatorname{ch}(t)}
m_{c\rightarrow t}
\left(
\mathbf a_{B_c\cap B_t}
\right)
\Bigg].
\label{eq:jt_message}
\end{align}
At the root bag $t_0$, the optimal objective is
\begin{equation}
\min_{\mathbf a_{B_{t_0}}}
\left[
\Psi_{t_0}(\mathbf a_{t_0})
+
\sum_{c\in\operatorname{ch}(t_0)}
m_{c\rightarrow t_0}
\left(
\mathbf a_{B_c\cap B_{t_0}}
\right)
\right].
\label{eq:jt_root}
\end{equation}
Backtracking through the minimizing assignments recovers a globally optimal workflow allocation.

Let
\[
M_{\max}
=
\max_{v\in\mathcal V_T}|\mathcal M_v|.
\]
For a supplied tree decomposition of width $w$, the running time is bounded by
\begin{equation}
O\left(
|\mathcal I|M_{\max}^{w+1}
\right)
\end{equation}
up to polynomial factors for evaluating local potentials and messages \citep{cygan2015parameterized}. For fixed treewidth, the dynamic program runs in polynomial time; it is fixed-parameter tractable when parameterized jointly by $w$ and $M_{\max}$.

With an available low-width decomposition, this junction-tree formulation provides an exact theoretical allocation algorithm. Its complexity is parameterized by the supplied decomposition width. The reported experimental pipeline uses the MILP in Section~\ref{sec:exact_allocation}.

\section{Synthetic Market and Workflow Configurations}
\label{app:market_config}

All four workflow families are hand-constructed finite DAGs. Nominal token values below are reported in thousands of tokens for readability; the solver converts them to millions of tokens before applying per-million-token costs. Unless otherwise stated, nominal values are sampled using Eq.~\eqref{eq:token_lognormal} below.

\subsection{Cost and Token Sampling}

We simulate nine providers in three capability tiers, with synthetic base-cost anchors following a $1{:}5{:}25$ progression. This progression provides a controlled separation between provider tiers. In the synthetic S1--S4 reference configuration,
\begin{equation}
c_i=C_i^{\mathrm{base}}S_i,\qquad S_i\sim\operatorname{Uniform}(0.1,1.0),
\label{eq:main_cost_draw}
\end{equation}
so provider $i$ has support
\begin{equation}
[a_i,d_i]=[0.1C_i^{\mathrm{base}},C_i^{\mathrm{base}}],
\qquad \phi_i(c)=2c-a_i.
\label{eq:main_uniform_support}
\end{equation}
We set $\rho_i=1/3$ for every provider, use $\kappa_{u,v}=0.1$ on cache-eligible edges, and set $\kappa_{u,v}=1$ on non-cacheable edges. The base-cost anchors remain fixed throughout this configuration.

For each nominal token quantity $\bar L$, we sample
\begin{equation}
L\sim\operatorname{LogNormal}\left(\log\bar L-\frac{\sigma^2}{2},\sigma^2\right),\qquad \sigma=0.3.
\label{eq:token_lognormal}
\end{equation}
The S2 source context is sampled separately from the configured truncated-Pareto distribution on $5\mathrm{k}$--$100\mathrm{k}$ tokens.

\subsection{Provider Tiers}

A provider in capability tier $t$ can execute tasks of difficulty at most $t$. In the workflow diagrams, D1, D2, and D3 denote public task-difficulty tiers 1, 2, and 3; ``Diff-1'' likewise means difficulty tier 1.

\begin{table}[h!]
    \centering
    \small
    \begin{tabular}{lcc}
    \toprule
    \textbf{Tier} & \textbf{Base-cost anchors} & \textbf{Providers}\\
    \midrule
    Tier 1 & \$0.10, \$0.20, \$0.30 & 3\\
    Tier 2 & \$0.50, \$1.00, \$1.50 & 3\\
    Tier 3 & \$2.50, \$5.00, \$7.50 & 3\\
    \bottomrule
    \end{tabular}
    \caption{Synthetic base-cost anchors in dollars per million standardized
    output-workload units.}
    \label{tab:provider_tiers}
\end{table}

\begin{table*}[t]
\centering
\small
\begin{tabular}{llccc}
\toprule
\textbf{Setting}
&
\textbf{Role}
&
\textbf{Count}
&
\textbf{Difficulty}
&
\textbf{Nominal output (k)}
\\
\midrule
S1
& Planner
& 5
& 1
& 300 each
\\
& Writers
& 15 (3 per stage)
& 2
& 40 each
\\
& Aggregator
& 5
& 1
& 50 each
\\
& Final Editor
& 1
& 1
& 20
\\
\midrule
S2
& Scanner / Analyzer / Planner
& 1 each
& 1 / 2 / 3
& 3 / 5 / 8
\\
& SpecGenerator
& 1
& 1
& 350
\\
& Modules
& 10
& $1\times3,\;7\times2,\;2\times1$
& 25 each
\\
& File tasks
& 30
& $10\times3,\;20\times1$
& 15 each
\\
& Merger / Tester
& 1 each
& 1 / 2
& 15 / 8
\\
\midrule
S3
& Architect / SpecWriter
& 1 each
& 2 / 1
& 10 / 300
\\
& Frontend / Backend / Interactive Lead
& 1 each
& 2 / 2 / 2
& 50 / 40 / 30
\\
& Frontend developers
& 5
& $3\times1,\;2\times2$
& 15 each
\\
& Backend developers
& 4
& $1\times1,\;3\times2$
& 20 each
\\
& Interactive developers
& 3
& $1\times2,\;2\times3$
& 15 each
\\
\midrule
S4
& Expert Assessment / Case Retriever
& 1 each
& 3 / 1
& 5 / 400
\\
& Analysts
& 5
& $3\times1,\;2\times2$
& 40, 35, 30, 20, 18
\\
& Aggregator
& 1
& 1
& 80
\\
& Strategists
& 3
& 2
& 20 each
\\
& Expert Synthesis
& 1
& 3
& 10
\\
\bottomrule
\end{tabular}
\caption{
Nominal synthetic workflow configurations. S1 repeats the Planner--Writers--Aggregator stage five times. In S2, the Scanner root context is sampled from the configured truncated-Pareto range, and the Merger receives all 30 file outputs, giving approximately 450k nominal fan-in tokens. The S3 Architect root input has 5k nominal tokens. The S4 Expert Assessment root input has 50k nominal tokens.
}
\label{tab:workflow_configurations}
\end{table*}

\tikzset{
    wfnode/.style={
        rectangle,
        rounded corners=3pt,
        draw=black!60,
        thick,
        align=center,
        font=\sffamily\scriptsize,
        minimum height=0.72cm,
        minimum width=1.75cm
    },
    d1/.style={wfnode,fill=cyan!15,draw=cyan!80!black},
    d2/.style={wfnode,fill=orange!15,draw=orange!80!black},
    d3/.style={wfnode,fill=red!15,draw=red!80!black},
    flow/.style={->,>=Stealth,thick,draw=gray!70},
    heavyflow/.style={
        ->,
        >=Stealth,
        line width=2.2pt,
        draw=blue!60!black
    },
    omission/.style={
        align=center,
        font=\sffamily\scriptsize,
        text=gray!75
    }
}

\begin{figure*}[t]
\centering
\begin{tikzpicture}[node distance=0.85cm and 1.0cm]
    \node[d1] (p1) {Planner 1\\D1, 300k};
    \node[d2,below=of p1] (w1) {3 Writers\\D2, 40k each};
    \node[d1,below=of w1] (a1) {Aggregator 1\\D1, 50k};

    \node[omission,below=0.55cm of a1] (dots)
    {$\vdots$\\Stages 2--4};

    \node[d1,below=0.55cm of dots] (p5) {Planner 5\\D1, 300k};
    \node[d2,below=of p5] (w5) {3 Writers\\D2, 40k each};
    \node[d1,below=of w5] (a5) {Aggregator 5\\D1, 50k};
    \node[d1,below=of a5] (final) {Final Editor\\D1, 20k};

    \draw[heavyflow] (p1) -- (w1);
    \draw[flow] (w1) -- (a1);
    \draw[flow] (a1) -- (dots);
    \draw[flow] (dots) -- (p5);
    \draw[heavyflow] (p5) -- (w5);
    \draw[flow] (w5) -- (a5);
    \draw[flow] (a5) -- (final);
\end{tikzpicture}
\caption{
\textbf{S1: Repeated Broadcasters.} Five repeated stages each contain one Diff-1 planner broadcasting a 300k-token plan to three Diff-2 writers, followed by a Diff-1 aggregator. The grouped writer boxes represent three parallel nodes per stage; stages 2--4 are omitted visually but are instantiated in the generator.
}
\Description{Workflow diagram for S1. A planner fans out to three writers, whose outputs enter an aggregator; this motif repeats for five stages and ends at a final editor.}
\label{fig:topo_s1}
\end{figure*}
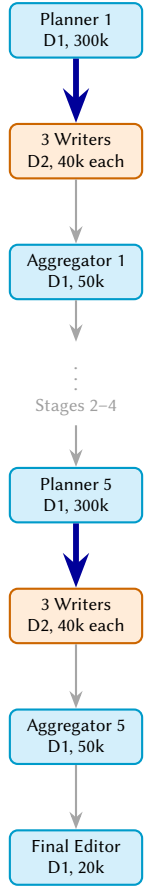

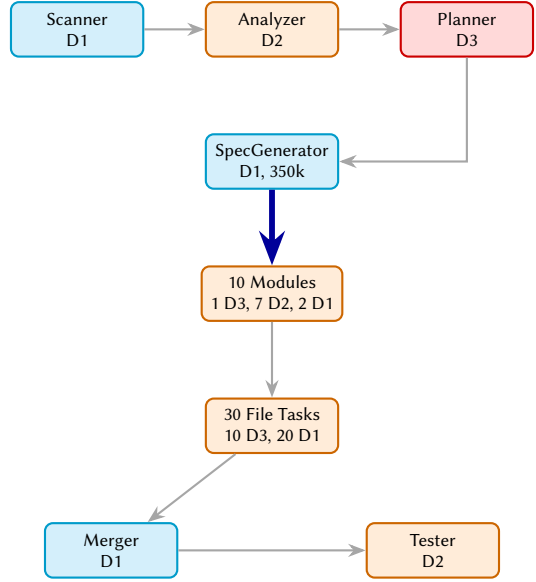
\begin{figure*}[t]
\centering
\begin{tikzpicture}[node distance=0.85cm and 0.8cm]
    \node[d1] (scanner) {Scanner\\D1};
    \node[d2,right=of scanner] (analyzer) {Analyzer\\D2};
    \node[d3,right=of analyzer] (planner) {Planner\\D3};

    \node[d1,below=1.0cm of analyzer] (spec)
    {SpecGenerator\\D1, 350k};

    \node[d2,below=1.0cm of spec] (modules)
    {10 Modules\\1 D3, 7 D2, 2 D1};

    \node[d2,below=1.0cm of modules] (files)
    {30 File Tasks\\10 D3, 20 D1};

    \node[d1,below left=0.9cm and 0.35cm of files] (merger)
    {Merger\\D1};
    \node[d2,below right=0.9cm and 0.35cm of files] (tester)
    {Tester\\D2};

    \draw[flow] (scanner) -- (analyzer);
    \draw[flow] (analyzer) -- (planner);
    \draw[flow] (planner) |- (spec);
    \draw[heavyflow] (spec) -- (modules);
    \draw[flow] (modules) -- (files);
    \draw[flow] (files) -- (merger);
    \draw[flow] (merger) -- (tester);
\end{tikzpicture}
\caption{
\textbf{S2: Nested Fan-Out.} A Diff-1 SpecGenerator broadcasts a 350k-token specification to ten modules, which decompose into thirty file tasks. All thirty file outputs feed the Merger, producing approximately 450k nominal fan-in tokens.
}
\Description{Workflow diagram for S2. Scanner, analyzer, and planner nodes lead to a specification generator, which fans out to ten modules and thirty file tasks before a merger and tester.}
\label{fig:topo_s2}
\end{figure*}

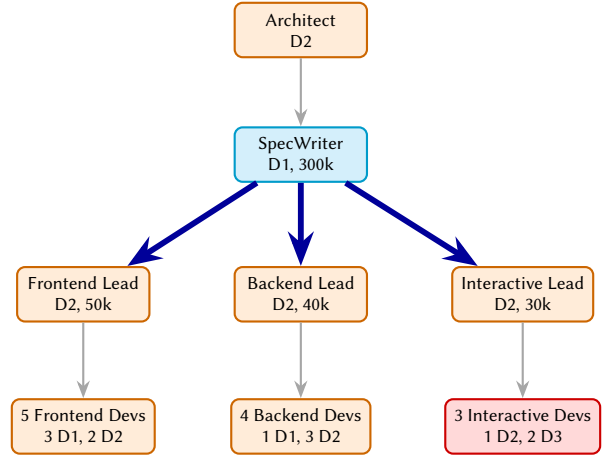
\begin{figure*}[t]
\centering
\begin{tikzpicture}[node distance=0.9cm and 1.0cm]
    \node[d2] (architect) {Architect\\D2};
    \node[d1,below=of architect] (spec)
    {SpecWriter\\D1, 300k};

    \node[d2,below=1.1cm of spec] (backend)
    {Backend Lead\\D2, 40k};
    \node[d2,left=1.1cm of backend] (frontend)
    {Frontend Lead\\D2, 50k};
    \node[d2,right=1.1cm of backend] (interactive)
    {Interactive Lead\\D2, 30k};

    \node[d2,below=1.0cm of frontend] (fdev)
    {5 Frontend Devs\\3 D1, 2 D2};
    \node[d2,below=1.0cm of backend] (bdev)
    {4 Backend Devs\\1 D1, 3 D2};
    \node[d3,below=1.0cm of interactive] (idev)
    {3 Interactive Devs\\1 D2, 2 D3};

    \draw[flow] (architect) -- (spec);
    \draw[heavyflow] (spec) -- (frontend);
    \draw[heavyflow] (spec) -- (backend);
    \draw[heavyflow] (spec) -- (interactive);
    \draw[flow] (frontend) -- (fdev);
    \draw[flow] (backend) -- (bdev);
    \draw[flow] (interactive) -- (idev);
\end{tikzpicture}
\caption{
\textbf{S3: Two-Level Team Fan-Out.} A Diff-1 SpecWriter broadcasts a 300k-token specification to three Diff-2 team leads, which route work to twelve developers with heterogeneous capability requirements.
}
\Description{Workflow diagram for S3. An architect and specification writer feed three parallel team leads for frontend, backend, and interactive work; each lead fans out to a group of developers.}
\label{fig:topo_s3}
\end{figure*}

\begin{figure*}[t]
\centering
\begin{tikzpicture}[node distance=0.9cm and 1.0cm]
    \node[d3] (expert) {Expert Assessment\\D3};
    \node[d1,below=of expert] (retriever)
    {Case Retriever\\D1, 400k};

    \node[d2,below=1.1cm of retriever] (analysts)
    {5 Analysts\\3 D1, 2 D2};

    \node[d1,below=1.0cm of analysts] (aggregator)
    {Aggregator\\D1, 80k};

    \node[d2,below=1.0cm of aggregator] (strategists)
    {3 Strategists\\D2, 20k each};

    \node[d3,below=1.0cm of strategists] (final)
    {Expert Synthesis\\D3, 10k};

    \draw[flow] (expert) -- (retriever);
    \draw[heavyflow] (retriever) -- (analysts);
    \draw[flow] (analysts) -- (aggregator);
    \draw[heavyflow] (aggregator) -- (strategists);
    \draw[flow] (strategists) -- (final);
\end{tikzpicture}
\caption{
\textbf{S4: Hourglass.}
A Diff-1 Case Retriever broadcasts a 400k-token payload to five analysts. Their outputs contract at a Diff-1 aggregator, expand to three Diff-2 strategists, and reconverge at the final Diff-3 synthesis node.
}
\Description{Workflow diagram for S4. Expert assessment and retrieval fan out to five analysts, contract at an aggregator, fan out again to three strategists, and reconverge at expert synthesis.}
\label{fig:topo_s4}
\end{figure*}
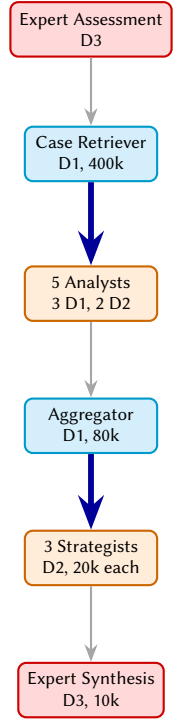

\section{Additional Experimental Results}
\label{app:additional_results}

\subsection{Synthetic S1--S4 Uniform-Prior Reference}
\label{app:uniform_reference}

\input{tables/calibrated_uniform_reference.tex}

The synthetic reference evaluates the four richer S1--S4 DAG families under provider-specific uniform priors. \textsc{Star} lowers sample-mean payment relative to Raw-DSIC by \(1.44\%\), \(0.36\%\), \(0.91\%\), and \(0.92\%\), respectively; the paired interval for S2 includes zero. This is a useful boundary case: when the raw and virtual score trade-offs are closely aligned, \textsc{Star} and Raw-DSIC often select similar allocations, so the additional payment benefit of virtual-cost optimization is modest.

\subsection{Heterogeneous-Prior Stress Configuration}
\label{app:stress_config}

Table~\ref{tab:prior_profile_parameters} specifies the fixed provider cost ranges and prior shapes. Experimental provider identifiers are 0--4; capability tier 2 can also execute tier-1 tasks. The structural cells are specified below.

\input{tables/prior_profile_parameters.tex}

\input{tables/structural_stress_configuration.tex}

The uniform, mild, and strong profiles use the same frozen structural cells. Every pre-specified cell is retained, including weak or nonsignificant effects. The documented provider-to-prior assignment is fixed across these cells, making the comparison reproducible for each specified market profile.

\subsection{Complete Heterogeneous-Prior Results}
\label{app:stress_results}

\input{tables/heterogeneous_prior_complete.tex}

For the representative degree-100 high-payload cell with \(\rho=1/3\), the two distinct effects are visible in the same comparison:
\begin{table*}[t]
\centering
\small
\begin{tabular}{lcc}
\toprule
\textbf{Prior profile} & \textbf{Raw graph-aware vs. Myopic execution reduction} & \textbf{\textsc{Star} vs. Raw-DSIC payment reduction}\\
\midrule
Uniform & \(91.11\%\,[90.63,91.48]\) & \(1.49\%\;[-0.39,3.58]\)\\
Heterogeneous mild & \(91.19\%\,[90.67,91.50]\) & \(10.23\%\;[3.44,16.73]\)\\
Heterogeneous strong & \(90.20\%\,[88.91,91.03]\) & \(31.48\%\;[22.56,40.73]\)\\
\bottomrule
\end{tabular}
\caption{Representative degree-100 high-payload cell at \(\rho=1/3\). The middle column isolates topology awareness, whereas the right column compares two graph-aware DSIC mechanisms and isolates the additional effect of virtual-cost scoring.}
\label{tab:representative_stress_cell}
\end{table*}

The graph-aware execution reduction remains near \(90\%\) across the three prior profiles, whereas the payment advantage varies strongly with prior shape. Thus, graph coupling and prior heterogeneity affect different comparisons.

\subsection{Complete Prior-Shape Continuum}
\label{app:prior_continuum}

\input{tables/prior_continuum_complete.tex}

\begin{figure*}[t]
\centering
\includegraphics[width=\textwidth]{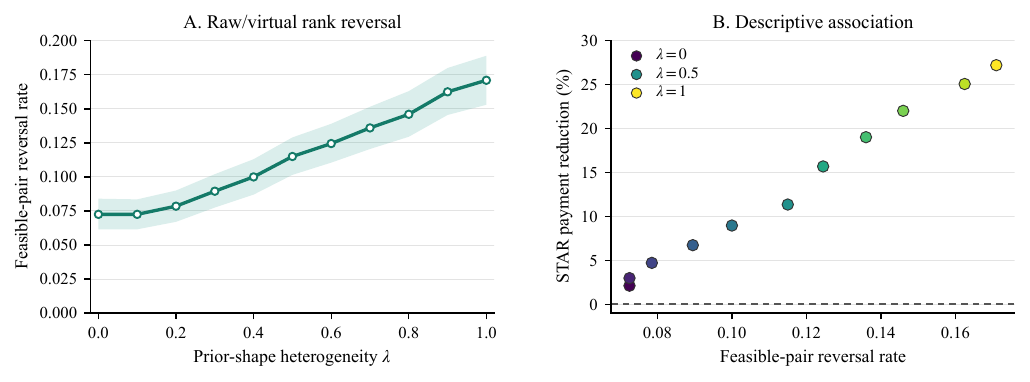}
\caption{\textbf{Prior-continuum diagnostics for degree 100.} The left panel reports the feasible-provider raw/virtual rank-reversal rate across \(\lambda\); the right panel summarizes its descriptive association with \textsc{Star}'s payment reduction along the fixed synthetic path.}
\Description{Two-panel diagnostic plot. The left panel shows rank-reversal rate across the prior-shape continuum, and the right panel plots payment reduction against rank-reversal rate.}
\label{fig:prior_continuum_diagnostics}
\end{figure*}

\input{tables/prior_continuum_degree_scaling.tex}

The normalized curves coincide across degrees 10, 20, and 100 because the frozen construction replicates identical consumer branches while preserving the set of competing provider pairs. Degree increases absolute payment savings. The mild profile serves as a separate off-path reference: its least-squares log-\(k\) projection is \(\lambda=0.338\), with a non-negligible residual.

\subsection{Structural Negative Controls}
\label{app:structural_controls}

Table~\ref{tab:structural_controls} in Section~\ref{sec:topology_controls} reports all three structural controls. The no-cache control makes graph-aware and myopic allocations agree row by row. The payload controls therefore support the mechanism in Proposition~\ref{prop:broadcaster_threshold}: fan-out creates a large topology advantage only when each repeated edge carries substantial avoidable fresh-input workload.

\subsection{Support-Width Heterogeneity as a Boundary Case}
\label{app:support_width}

\input{tables/region_support_width.tex}

The region-inspired experiment varies synthetic support-width profiles while holding normalized provider means fixed, allowing aggregate dispersion to change with the treatment. In the tested degree-100 and S4 cells, the additional treatment-control payment reduction is modest and its 95\% interval includes zero. These paired estimates characterize the payment effect and uncertainty for the documented synthetic profiles, complementing the prior-shape comparisons above.

\subsection{Prior Misspecification and Runtime Details}
\label{app:runtime_details}

\paragraph{Prior-estimation-error magnitude grid (E2).}
The experiment uses the original five provider cost ranges, eligibility rules, degree-10 fan-out graph, $\rho=1/3$, and cache semantics. The uniform profile has $k_i=1$ for all providers; the heterogeneous power-prior profile uses $k=(0.5,2,0.2,5,1)$, the same provider-specific shapes used in Table~\ref{tab:e2_misspecification_main}. Appendix~\ref{app:stress_config} gives their cost ranges. For each profile, we evaluate eight scoring conditions: correct-prior \textsc{Star}, six signed shape perturbations $\widehat k_i=k_i e^{\delta r_i}$ with $r=(1,-1,0,1,-1)$ and $\delta\in\{-0.2,-0.1,-0.05,+0.05,+0.1,+0.2\}$, and Raw-DSIC. These controlled parameter errors leave each provider's true cost and truthful report unchanged.

Before execution, we fix 1,000 new draw identifiers (30,000--30,999) per profile and retain every draw and signed perturbation. Each draw's latent quantiles are shared across the two true profiles, and within a profile all scoring conditions share the identical true cost vector. The shapes obey $\widehat k_i/k_i\in[e^{-m},e^{m}]$ at magnitude $m=|\delta|$. For $m=0.05,0.1,0.2$, the maximum decreases are $4.88\%$, $9.52\%$, and $18.13\%$, and the maximum increases are $5.13\%$, $10.52\%$, and $22.14\%$, respectively. The fixed direction permits a matched comparison of both signs as the error magnitude grows.

Table~\ref{tab:e2_prior_error_grid_appendix} gives all condition-level payment decompositions and changes from correct-prior routing. All 95\% intervals use 2,000 resamples of the 1,000 paired draw identifiers, with scoring conditions resampled together. The new cohort is analyzed separately from earlier runs; its means are not pooled with earlier draws.

\input{tables/e2_prior_error_grid_appendix.tex}

\paragraph{Optimization runtime (E1).}
These central processing unit (CPU) measurements time the task-assignment and payment optimizations required to run \textsc{Star} on a supplied workflow and provider reports. Synthetic graphs provide controlled inputs to those same computations. The reported mechanism time includes constructing and solving the allocation problems, report-independent tie-breaking, and every active provider's payment computation; graph generation is outside this timer.

On one Apple M2 Pro processor, all 36 chain/sparse-DAG instances complete within the time limits. We test 8, 32, and 64 task nodes, with 3 or 5 providers and three seeds. The largest complete-mechanism time is \(0.767\) seconds. Peak sampled memory is \(268.9\) mebibytes (MiB), measured as the summed resident set size (RSS, physical memory occupied) of the Python worker and all CBC solver processes it launches, including their subprocesses. Memory is sampled every 20 milliseconds.

CBC uses one thread, zero requested relative and absolute optimality gap, and a 30-second limit for each allocation optimization. Each complete mechanism has a 180-second limit. Table~\ref{tab:e1_scaling_main} summarizes the timings by workflow size, and Table~\ref{tab:e1_scaling_appendix} gives the graph-family and provider-count breakdown.

\input{tables/e1_scaling_main.tex}

\input{tables/e1_scaling_appendix.tex}

\subsection{Experiment Coverage and Validation}
\label{app:coverage_validation}

The evaluation includes every pre-specified synthetic structural-stress and prior-continuum cell. E1 completes 36/36 instances, and E2 includes all 16,000 pre-specified condition-level results across both true profiles and all eight scoring conditions per profile. Every allocation optimization reports status \texttt{Optimal} with zero reported gap; the E2 minimum truthful provider utility is zero and payment-decomposition errors are at numerical precision. In every checked instance, when providers report their actual unit costs, Raw-DSIC assigns each task to exactly the same provider as the execution-cost-minimizing oracle.

\clearpage
\section{Disclosure of AI Assistance}
\label{app:ai_use}

OpenAI Codex assisted with language editing, \LaTeX{} formatting, and implementing, running, and checking experiment code under the authors' direction. The authors are responsible for the final manuscript, experiments, and references.

\end{document}

%% file: tables/heterogeneous_prior_summary.tex
\begin{table}[t]
\centering
\setlength{\tabcolsep}{5pt}

\begin{tabular}{lcc}
\toprule
Prior profile & Reduction range & Positive cells \\
\midrule
Uniform & 1.49\%--3.34\% & 2/24 \\
Heterogeneous mild & 10.23\%--10.30\% & 24/24 \\
Heterogeneous strong & 21.99\%--31.48\% & 24/24 \\
\bottomrule
\end{tabular}
\caption{\textsc{Star} payment reductions across prior profiles. Ranges are the minimum and maximum point estimates over all 24 pre-specified structural cells; all cells are retained. A positive cell has a paired-bootstrap 95\% interval strictly above zero.}
\label{tab:heterogeneous_prior_summary}
\end{table}

%% file: tables/e2_misspecification_main.tex
\begin{table*}[t]
\centering
\normalsize
\setlength{\tabcolsep}{4pt}
\begin{tabular}{lrrrr}
\toprule
\shortstack{True cost\\distribution} & \shortstack{Correct-prior \textsc{Star}\\reduction (95\% CI)} & $|\delta|$ & \shortstack{Negative-direction error\\$\delta=-|\delta|$ (95\% CI)} & \shortstack{Positive-direction error\\$\delta=+|\delta|$ (95\% CI)}\\
\midrule
\multirow{3}{*}{Uniform} & \multirow{3}{*}{0.79\% [0.31,1.25]} & 0.05 & 0.85\% [0.28,1.38] & 0.72\% [0.26,1.15]\\
 &  & 0.1 & 0.71\% [0.07,1.32] & 0.56\% [0.10,0.97]\\
 &  & 0.2 & 0.34\% [-0.54,1.16] & 0.20\% [-0.31,0.69]\\
\midrule
\multirow{3}{*}{\shortstack[l]{Heterogeneous\\power priors}} & \multirow{3}{*}{30.04\% [27.88,32.21]} & 0.05 & 30.01\% [27.84,32.14] & 29.98\% [27.79,32.09]\\
 &  & 0.1 & 30.15\% [28.02,32.29] & 29.92\% [27.76,32.03]\\
 &  & 0.2 & 30.16\% [28.06,32.27] & 30.13\% [27.93,32.23]\\
\bottomrule
\end{tabular}
\caption{User-payment savings (\%) under prior-estimation errors. All three result columns and their bracketed paired 95\% bootstrap confidence intervals show the user's total-payment savings relative to Raw-DSIC in percent: $100(1-\overline P_{\textsc{Star}}/\overline P_{\mathrm{Raw}})\%$. Positive values mean lower user payments. The perturbation magnitude $|\delta|$ is dimensionless. Only assumed shapes change: $\widehat k_i=k_i e^{\delta r_i}$, with fixed $r=(1,-1,0,1,-1)$. Uniform uses $k_i=1$ for every provider; heterogeneous power priors use $k=(0.5,2,0.2,5,1)$. Each three-row group shares 1,000 cost draws and correct-prior and Raw-DSIC baselines; the merged correct-prior entry applies to all three magnitudes. True distributions, graph, and cost ranges stay fixed. Appendix~\ref{app:runtime_details} provides payment details.}
\label{tab:e2_misspecification_main}
\end{table*}

%% file: tables/structural_controls.tex
\begin{table}[t]
\centering
\normalsize
\setlength{\tabcolsep}{2pt}

\begin{tabular}{@{}p{0.49\columnwidth}cc@{}}
\toprule
Control & \shortstack{Execution\\cost\\reduction} & Paired 95\% CI \\
\midrule
Low payload (ratio 1) & 2.13\% & [1.16\%, 3.24\%] \\
Higher payload (ratio 16) & 67.47\% & [66.12\%, 68.48\%] \\
No cache advantage ($\kappa=1$) & 0.00\% & [0.00\%, 0.00\%] \\
\bottomrule
\end{tabular}
\caption{Degree-100 fan-out controls ($\rho=1/3$, uniform priors). Ratio is transferred context per edge divided by each child's $20,000$-token output: ratios 1 and 16 transfer $20,000$ and $320,000$ tokens, respectively, with $\kappa=0.1$; the no-cache row transfers $1,000,000$ tokens with $\kappa=1$. Hub output is $1,000,000$ tokens throughout. Table~\ref{tab:structural_stress_configuration} instead uses ratio 50 and $\kappa=0$. Reductions compare graph-aware Raw-DSIC execution cost with Myopic; intervals are paired-bootstrap 95\% confidence intervals. The $\kappa=1$ zeros are exact, not rounded: without reuse discounts and with common $\rho$, Raw-DSIC and Myopic attain the same execution cost.}
\label{tab:structural_controls}
\end{table}

%% file: tables/calibrated_uniform_reference.tex
\begin{table}[t]
\centering
\small
\begin{tabular}{lcc}
\toprule
Workflow & Payment reduction & Paired 95\% CI \\
\midrule
S1 & 1.44\% & [0.14\%, 2.72\%] \\
S2 & 0.36\% & [-0.39\%, 1.13\%] \\
S3 & 0.91\% & [0.07\%, 1.79\%] \\
S4 & 0.92\% & [0.11\%, 1.71\%] \\
\bottomrule
\end{tabular}
\caption{Synthetic S1--S4 uniform-prior boundary reference. Intervals are paired-bootstrap 95\% confidence intervals for \textsc{Star} relative to graph-aware Raw-DSIC.}
\label{tab:calibrated_uniform_reference}
\end{table}

%% file: tables/prior_profile_parameters.tex
\begin{table*}[t]
\centering
\normalsize
\begin{tabular}{ccccccc}
\toprule
Provider ID & \shortstack{Capability\\tier} & \multicolumn{2}{c}{\shortstack{Unit-cost range\\(\$/million work units)}} & \shortstack{Uniform\\$k_i$} & \shortstack{Mild\\$k_i$} & \shortstack{Strong\\$k_i$} \\
\cmidrule(lr){3-4}
 & & $a_i$ & $d_i$ & & & \\
\midrule
0 & 1 & 0.05 & 0.50 & 1.0 & 0.5 & 0.5 \\
1 & 1 & 0.10 & 0.80 & 1.0 & 2.0 & 2.0 \\
2 & 2 & 0.25 & 1.40 & 1.0 & 0.5 & 0.2 \\
3 & 2 & 0.35 & 1.10 & 1.0 & 1.0 & 5.0 \\
4 & 2 & 0.45 & 1.25 & 1.0 & 2.0 & 1.0 \\
\bottomrule
\end{tabular}
\caption{Fixed provider cost ranges and distribution shapes for the structural-prior stress suite. Work units are the standardized workloads in Eq.~\eqref{eq:provider_workload}. The cost ranges stay fixed across profiles; the listed shape parameters determine the virtual costs $\phi_i(c)=(1+1/k_i)c-a_i/k_i$.}
\label{tab:prior_profile_parameters}
\end{table*}

%% file: tables/structural_stress_configuration.tex
\begin{table}[p]
\centering
\small

\begin{tabular}{clrrr}
\toprule
Cell & Topology & Degree & $\rho$ & Paired trials \\
\midrule
C01 & Fanout & 1 & $1/3$ & 50 \\
C02 & Fanout & 1 & 1.00 & 50 \\
C03 & Fanout & 10 & $1/3$ & 50 \\
C04 & Fanout & 10 & 1.00 & 50 \\
C05 & Fanout & 30 & $1/3$ & 50 \\
C06 & Fanout & 30 & 1.00 & 50 \\
C07 & Fanout & 100 & $1/3$ & 50 \\
C08 & Fanout & 100 & 1.00 & 50 \\
C09 & Fanin & 1 & $1/3$ & 50 \\
C10 & Fanin & 1 & 1.00 & 50 \\
C11 & Fanin & 10 & $1/3$ & 50 \\
C12 & Fanin & 10 & 1.00 & 50 \\
C13 & Fanin & 30 & $1/3$ & 50 \\
C14 & Fanin & 30 & 1.00 & 50 \\
C15 & Fanin & 100 & $1/3$ & 50 \\
C16 & Fanin & 100 & 1.00 & 50 \\
C17 & Hourglass & 1 & $1/3$ & 50 \\
C18 & Hourglass & 1 & 1.00 & 50 \\
C19 & Hourglass & 10 & $1/3$ & 50 \\
C20 & Hourglass & 10 & 1.00 & 50 \\
C21 & Hourglass & 30 & $1/3$ & 50 \\
C22 & Hourglass & 30 & 1.00 & 50 \\
C23 & Hourglass & 100 & $1/3$ & 50 \\
C24 & Hourglass & 100 & 1.00 & 50 \\
\bottomrule
\end{tabular}
\caption{The 24 frozen structural cells used separately under each of the uniform, mild, and strong prior profiles. The fixed settings are $\kappa=0.0$, a transferred payload of $1,000,000$ tokens, a generated hub payload of $1,000,000$ tokens, and child outputs of $20,000$ tokens.}
\label{tab:structural_stress_configuration}
\end{table}

%% file: tables/heterogeneous_prior_complete.tex
\begin{table*}[p]
\centering
\small
\setlength{\tabcolsep}{3.3pt}

\begin{tabular}{lrrcccc}
\toprule
Topology & Degree & $\rho$ & \textsc{Star}--Raw payment & Paired 95\% CI & Raw--Myopic execution & Paired 95\% CI \\
\midrule
Fanout & 1 & $1/3$ & 3.34\% & [0.52\%, 6.07\%] & 1.77\% & [0.57\%, 3.31\%] \\
Fanout & 1 & 1.00 & 1.49\% & [-0.36\%, 3.41\%] & 24.15\% & [20.89\%, 27.28\%] \\
Fanout & 10 & $1/3$ & 1.49\% & [-0.44\%, 3.49\%] & 67.79\% & [66.31\%, 68.86\%] \\
Fanout & 10 & 1.00 & 1.49\% & [-0.39\%, 3.48\%] & 87.84\% & [87.16\%, 88.34\%] \\
Fanout & 30 & $1/3$ & 1.49\% & [-0.46\%, 3.48\%] & 84.60\% & [83.79\%, 85.19\%] \\
Fanout & 30 & 1.00 & 1.49\% & [-0.37\%, 3.41\%] & 94.47\% & [94.12\%, 94.71\%] \\
Fanout & 100 & $1/3$ & 1.49\% & [-0.39\%, 3.58\%] & 91.11\% & [90.63\%, 91.48\%] \\
Fanout & 100 & 1.00 & 1.49\% & [-0.38\%, 3.41\%] & 96.88\% & [96.68\%, 97.02\%] \\
Fanin & 1 & $1/3$ & 1.49\% & [-0.47\%, 3.50\%] & 9.07\% & [7.47\%, 10.74\%] \\
Fanin & 1 & 1.00 & 1.49\% & [-0.53\%, 3.50\%] & 24.15\% & [20.77\%, 27.45\%] \\
Fanin & 10 & $1/3$ & 1.49\% & [-0.49\%, 3.41\%] & 10.09\% & [8.56\%, 11.62\%] \\
Fanin & 10 & 1.00 & 1.49\% & [-0.52\%, 3.42\%] & 25.07\% & [21.80\%, 28.15\%] \\
Fanin & 30 & $1/3$ & 1.49\% & [-0.37\%, 3.53\%] & 10.17\% & [8.68\%, 11.76\%] \\
Fanin & 30 & 1.00 & 1.49\% & [-0.48\%, 3.47\%] & 25.14\% & [22.02\%, 28.34\%] \\
Fanin & 100 & $1/3$ & 1.49\% & [-0.49\%, 3.43\%] & 10.19\% & [8.68\%, 11.74\%] \\
Fanin & 100 & 1.00 & 1.49\% & [-0.55\%, 3.52\%] & 25.17\% & [21.90\%, 28.29\%] \\
Hourglass & 1 & $1/3$ & 2.31\% & [0.58\%, 4.11\%] & 6.28\% & [5.01\%, 7.65\%] \\
Hourglass & 1 & 1.00 & 1.49\% & [-0.54\%, 3.48\%] & 24.51\% & [21.19\%, 27.61\%] \\
Hourglass & 10 & $1/3$ & 1.49\% & [-0.43\%, 3.47\%] & 24.98\% & [23.36\%, 26.52\%] \\
Hourglass & 10 & 1.00 & 1.49\% & [-0.40\%, 3.54\%] & 52.57\% & [50.53\%, 54.19\%] \\
Hourglass & 30 & $1/3$ & 1.49\% & [-0.42\%, 3.42\%] & 28.30\% & [26.73\%, 29.65\%] \\
Hourglass & 30 & 1.00 & 1.49\% & [-0.50\%, 3.50\%] & 55.01\% & [53.12\%, 56.48\%] \\
Hourglass & 100 & $1/3$ & 1.49\% & [-0.40\%, 3.47\%] & 29.49\% & [28.00\%, 30.82\%] \\
Hourglass & 100 & 1.00 & 1.49\% & [-0.46\%, 3.44\%] & 55.88\% & [53.94\%, 57.45\%] \\
\bottomrule
\end{tabular}
\caption{Complete frozen structural-grid results for the uniform prior. Payment columns compare \textsc{Star} with graph-aware Raw-DSIC; execution columns compare graph-aware Raw with Myopic. All 24 pre-specified cells are retained, and intervals are paired-bootstrap 95\% confidence intervals.}
\label{tab:heterogeneous_prior_complete}
\end{table*}

\begin{table*}[p]
\centering
\small
\setlength{\tabcolsep}{3.3pt}

\begin{tabular}{lrrcccc}
\toprule
Topology & Degree & $\rho$ & \textsc{Star}--Raw payment & Paired 95\% CI & Raw--Myopic execution & Paired 95\% CI \\
\midrule
Fanout & 1 & $1/3$ & 10.30\% & [3.26\%, 16.70\%] & 2.27\% & [0.71\%, 4.17\%] \\
Fanout & 1 & 1.00 & 10.23\% & [3.42\%, 16.73\%] & 22.65\% & [18.73\%, 26.62\%] \\
Fanout & 10 & $1/3$ & 10.23\% & [3.26\%, 16.47\%] & 67.75\% & [66.33\%, 68.85\%] \\
Fanout & 10 & 1.00 & 10.23\% & [2.85\%, 16.82\%] & 87.92\% & [87.29\%, 88.35\%] \\
Fanout & 30 & $1/3$ & 10.23\% & [3.45\%, 16.98\%] & 84.68\% & [83.88\%, 85.21\%] \\
Fanout & 30 & 1.00 & 10.23\% & [3.25\%, 16.81\%] & 94.52\% & [94.22\%, 94.72\%] \\
Fanout & 100 & $1/3$ & 10.23\% & [3.44\%, 16.73\%] & 91.19\% & [90.67\%, 91.50\%] \\
Fanout & 100 & 1.00 & 10.23\% & [3.50\%, 16.60\%] & 96.91\% & [96.72\%, 97.02\%] \\
Fanin & 1 & $1/3$ & 10.23\% & [3.53\%, 16.63\%] & 8.31\% & [6.44\%, 10.26\%] \\
Fanin & 1 & 1.00 & 10.23\% & [3.21\%, 16.91\%] & 22.65\% & [18.64\%, 26.61\%] \\
Fanin & 10 & $1/3$ & 10.23\% & [3.27\%, 16.88\%] & 9.38\% & [7.61\%, 11.21\%] \\
Fanin & 10 & 1.00 & 10.23\% & [3.54\%, 16.95\%] & 23.61\% & [19.60\%, 27.38\%] \\
Fanin & 30 & $1/3$ & 10.23\% & [3.68\%, 16.96\%] & 9.47\% & [7.72\%, 11.24\%] \\
Fanin & 30 & 1.00 & 10.23\% & [3.28\%, 16.66\%] & 23.69\% & [19.99\%, 27.49\%] \\
Fanin & 100 & $1/3$ & 10.23\% & [3.73\%, 17.08\%] & 9.49\% & [7.73\%, 11.35\%] \\
Fanin & 100 & 1.00 & 10.23\% & [3.17\%, 16.94\%] & 23.71\% & [19.82\%, 27.42\%] \\
Hourglass & 1 & $1/3$ & 10.26\% & [4.23\%, 16.00\%] & 6.06\% & [4.49\%, 7.70\%] \\
Hourglass & 1 & 1.00 & 10.23\% & [3.21\%, 16.57\%] & 22.99\% & [19.09\%, 26.89\%] \\
Hourglass & 10 & $1/3$ & 10.23\% & [3.40\%, 16.92\%] & 24.51\% & [22.74\%, 26.19\%] \\
Hourglass & 10 & 1.00 & 10.23\% & [3.45\%, 16.79\%] & 52.16\% & [50.08\%, 54.01\%] \\
Hourglass & 30 & $1/3$ & 10.23\% & [3.23\%, 16.66\%] & 27.96\% & [26.39\%, 29.50\%] \\
Hourglass & 30 & 1.00 & 10.23\% & [3.27\%, 16.39\%] & 54.66\% & [52.74\%, 56.47\%] \\
Hourglass & 100 & $1/3$ & 10.23\% & [3.49\%, 16.92\%] & 29.20\% & [27.70\%, 30.64\%] \\
Hourglass & 100 & 1.00 & 10.23\% & [3.50\%, 16.69\%] & 55.55\% & [53.69\%, 57.16\%] \\
\bottomrule
\end{tabular}
\caption{Complete frozen structural-grid results for the heterogeneous mild prior. Payment columns compare \textsc{Star} with graph-aware Raw-DSIC; execution columns compare graph-aware Raw with Myopic. All 24 pre-specified cells are retained, and intervals are paired-bootstrap 95\% confidence intervals.}
\label{tab:heterogeneous_prior_complete_mild}
\end{table*}

\begin{table*}[p]
\centering
\small
\setlength{\tabcolsep}{3.3pt}

\begin{tabular}{lrrcccc}
\toprule
Topology & Degree & $\rho$ & \textsc{Star}--Raw payment & Paired 95\% CI & Raw--Myopic execution & Paired 95\% CI \\
\midrule
Fanout & 1 & $1/3$ & 21.99\% & [14.04\%, 29.82\%] & 2.52\% & [0.80\%, 4.58\%] \\
Fanout & 1 & 1.00 & 31.48\% & [21.92\%, 40.29\%] & 21.82\% & [17.46\%, 25.88\%] \\
Fanout & 10 & $1/3$ & 31.48\% & [21.96\%, 40.40\%] & 65.32\% & [62.29\%, 67.52\%] \\
Fanout & 10 & 1.00 & 31.48\% & [21.78\%, 40.66\%] & 86.63\% & [84.98\%, 87.72\%] \\
Fanout & 30 & $1/3$ & 31.48\% & [21.78\%, 40.41\%] & 83.13\% & [81.04\%, 84.43\%] \\
Fanout & 30 & 1.00 & 31.48\% & [21.60\%, 40.12\%] & 93.88\% & [93.05\%, 94.40\%] \\
Fanout & 100 & $1/3$ & 31.48\% & [22.56\%, 40.73\%] & 90.20\% & [88.91\%, 91.03\%] \\
Fanout & 100 & 1.00 & 31.48\% & [22.30\%, 40.32\%] & 96.53\% & [96.05\%, 96.84\%] \\
Fanin & 1 & $1/3$ & 31.48\% & [22.27\%, 40.39\%] & 8.02\% & [6.05\%, 9.96\%] \\
Fanin & 1 & 1.00 & 31.48\% & [21.98\%, 40.82\%] & 21.82\% & [17.44\%, 26.01\%] \\
Fanin & 10 & $1/3$ & 31.48\% & [21.82\%, 40.50\%] & 8.96\% & [7.09\%, 11.02\%] \\
Fanin & 10 & 1.00 & 31.48\% & [21.65\%, 40.27\%] & 22.70\% & [18.51\%, 26.69\%] \\
Fanin & 30 & $1/3$ & 31.48\% & [21.87\%, 40.61\%] & 9.03\% & [7.13\%, 10.98\%] \\
Fanin & 30 & 1.00 & 31.48\% & [21.48\%, 40.63\%] & 22.76\% & [18.55\%, 26.72\%] \\
Fanin & 100 & $1/3$ & 31.48\% & [21.84\%, 40.35\%] & 9.06\% & [7.19\%, 11.00\%] \\
Fanin & 100 & 1.00 & 31.48\% & [22.18\%, 40.90\%] & 22.78\% & [18.55\%, 26.99\%] \\
Hourglass & 1 & $1/3$ & 27.55\% & [18.85\%, 35.28\%] & 5.87\% & [4.17\%, 7.77\%] \\
Hourglass & 1 & 1.00 & 31.48\% & [21.52\%, 41.01\%] & 22.15\% & [17.71\%, 26.42\%] \\
Hourglass & 10 & $1/3$ & 31.48\% & [22.28\%, 40.38\%] & 22.85\% & [20.33\%, 25.12\%] \\
Hourglass & 10 & 1.00 & 31.48\% & [22.19\%, 40.72\%] & 49.70\% & [45.98\%, 52.66\%] \\
Hourglass & 30 & $1/3$ & 31.48\% & [21.86\%, 40.62\%] & 26.03\% & [23.33\%, 28.28\%] \\
Hourglass & 30 & 1.00 & 31.48\% & [21.73\%, 40.69\%] & 52.17\% & [48.63\%, 54.82\%] \\
Hourglass & 100 & $1/3$ & 31.48\% & [21.53\%, 40.47\%] & 27.17\% & [24.50\%, 29.44\%] \\
Hourglass & 100 & 1.00 & 31.48\% & [21.85\%, 40.53\%] & 53.05\% & [49.61\%, 55.75\%] \\
\bottomrule
\end{tabular}
\caption{Complete frozen structural-grid results for the heterogeneous strong prior. Payment columns compare \textsc{Star} with graph-aware Raw-DSIC; execution columns compare graph-aware Raw with Myopic. All 24 pre-specified cells are retained, and intervals are paired-bootstrap 95\% confidence intervals.}
\label{tab:heterogeneous_prior_complete_strong}
\end{table*}

%% file: tables/prior_continuum_complete.tex
\begin{table*}[t]
\centering

\begin{tabular}{ccccc}
\toprule
$\lambda$ & Payment reduction & Paired 95\% CI & Pair-reversal rate & Allocation-disagreement rate \\
\midrule
0.0 & 2.11\% & [1.11\%, 3.10\%] & 0.072 & 0.060 \\
0.1 & 2.97\% & [1.51\%, 4.44\%] & 0.072 & 0.090 \\
0.2 & 4.71\% & [2.58\%, 6.78\%] & 0.079 & 0.100 \\
0.3 & 6.73\% & [3.91\%, 9.52\%] & 0.089 & 0.140 \\
0.4 & 8.96\% & [5.79\%, 12.10\%] & 0.100 & 0.210 \\
0.5 & 11.35\% & [7.86\%, 14.93\%] & 0.115 & 0.250 \\
0.6 & 15.69\% & [11.87\%, 19.54\%] & 0.124 & 0.235 \\
0.7 & 19.02\% & [14.81\%, 23.22\%] & 0.136 & 0.220 \\
0.8 & 22.02\% & [17.44\%, 26.35\%] & 0.146 & 0.240 \\
0.9 & 25.08\% & [20.28\%, 29.65\%] & 0.163 & 0.255 \\
1.0 & 27.22\% & [22.41\%, 32.02\%] & 0.171 & 0.285 \\
\bottomrule
\end{tabular}
\caption{Complete degree-100 prior-shape continuum. Supports and high-payload graph semantics are fixed; the mild prior serves as a separate off-path reference. Rank reversal and allocation disagreement are reported as rates, and payment intervals are paired-bootstrap 95\% confidence intervals.}
\label{tab:prior_continuum_complete}
\end{table*}

%% file: tables/prior_continuum_degree_scaling.tex
\begin{table}[t]
\centering
\footnotesize
\setlength{\tabcolsep}{3pt}
\begin{tabular}{rcc}
\toprule
Degree & Absolute saving (\$) & Reduction \\
\midrule
10 & 0.276 & 27.22\% \\
20 & 0.322 & 27.22\% \\
100 & 0.687 & 27.22\% \\
\bottomrule
\end{tabular}
\caption{Degree scaling at $\lambda=1$. Absolute savings grow with replicated workload, while normalized reductions coincide under the frozen replication-scaled construction.}
\label{tab:prior_continuum_degree_scaling}
\end{table}

%% file: tables/region_support_width.tex
\begin{table*}[t]
\centering
\small
\setlength{\tabcolsep}{3.2pt}

\begin{tabular}{lrcccc}
\toprule
Structural cell & $\eta$ & Treatment [95\% CI] & Control [95\% CI] & Treatment--control [95\% CI] & CI $>0$ \\
\midrule
Degree-100 fan-out & 0.25 & 0.000\% [0.000\%, 0.000\%] & 0.000\% [0.000\%, 0.000\%] & 0.000\% [0.000\%, 0.000\%] & No \\
Degree-100 fan-out & 0.50 & 0.096\% [-1.718\%, 1.804\%] & -0.323\% [-2.167\%, 1.088\%] & 0.419\% [-0.972\%, 1.723\%] & No \\
Degree-100 fan-out & 1.00 & 2.538\% [-0.658\%, 5.790\%] & 1.658\% [-0.545\%, 4.082\%] & 0.879\% [-2.050\%, 3.744\%] & No \\
S4 cached input & 0.25 & 0.005\% [0.000\%, 0.012\%] & 0.000\% [0.000\%, 0.000\%] & 0.005\% [0.000\%, 0.012\%] & No \\
S4 cached input & 0.50 & 0.256\% [-0.919\%, 1.429\%] & -0.316\% [-1.393\%, 0.595\%] & 0.572\% [-0.387\%, 1.477\%] & No \\
S4 cached input & 1.00 & 3.385\% [0.973\%, 5.752\%] & 1.748\% [-0.232\%, 3.741\%] & 1.637\% [-0.271\%, 3.602\%] & No \\
\bottomrule
\end{tabular}
\caption{Region-inspired support-width contrasts. Treatment and control preserve provider means while varying support widths and aggregate dispersion. The parameter $\eta$ scales each cost range's half-width around its fixed mean: treatment half-widths are 20\%, 50\%, or 90\% of the mean, multiplied by $\eta$; control half-widths are uniformly $50\%\times\eta$. The S4 cell uses cached-input fraction $\kappa=0.1$. Each paired-bootstrap interval resamples $n=50$ seed blocks, retaining all six profile assignments together within each block. The pre-specified materiality assessment uses a 5-percentage-point threshold; the final column checks whether the treatment--control interval lies strictly above zero.}
\label{tab:region_support_width}
\end{table*}

%% file: tables/e2_prior_error_grid_appendix.tex
\begin{table*}[t]
\centering
\normalsize
\begin{tabular}{llrrrrr}
\toprule
True prior & Scoring condition & $P$ & $C_{\rm exec}$ & $R$ & \shortstack{$P-P_{\rm correct}$\\(95\% CI)} & \shortstack{Allocation\\disagreement}\\
\midrule
Uniform & Correct prior & 0.939 & 0.686 & 0.254 & +0.0000 [+0.0000,+0.0000] & 0.0\%\\
Uniform & $\delta=-0.2$ & 0.944 & 0.692 & 0.252 & +0.0043 [-0.0012,+0.0096] & 7.0\%\\
Uniform & $\delta=-0.1$ & 0.940 & 0.688 & 0.252 & +0.0008 [-0.0019,+0.0035] & 3.5\%\\
Uniform & $\delta=-0.05$ & 0.939 & 0.686 & 0.253 & -0.0005 [-0.0019,+0.0008] & 1.0\%\\
Uniform & $\delta=+0.05$ & 0.940 & 0.685 & 0.255 & +0.0007 [-0.0006,+0.0020] & 1.9\%\\
Uniform & $\delta=+0.1$ & 0.942 & 0.685 & 0.256 & +0.0023 [-0.0002,+0.0047] & 3.4\%\\
Uniform & $\delta=+0.2$ & 0.945 & 0.686 & 0.260 & +0.0056 [+0.0013,+0.0102] & 6.8\%\\
Uniform & Raw-DSIC & 0.947 & 0.682 & 0.265 & +0.0075 [+0.0029,+0.0119] & 8.9\%\\
\midrule
Heterogeneous & Correct prior & 0.694 & 0.564 & 0.130 & +0.0000 [+0.0000,+0.0000] & 0.0\%\\
Heterogeneous & $\delta=-0.2$ & 0.693 & 0.557 & 0.136 & -0.0012 [-0.0052,+0.0025] & 3.1\%\\
Heterogeneous & $\delta=-0.1$ & 0.693 & 0.560 & 0.133 & -0.0010 [-0.0042,+0.0018] & 1.7\%\\
Heterogeneous & $\delta=-0.05$ & 0.694 & 0.563 & 0.132 & +0.0003 [-0.0013,+0.0018] & 0.7\%\\
Heterogeneous & $\delta=+0.05$ & 0.695 & 0.565 & 0.129 & +0.0006 [-0.0009,+0.0027] & 0.5\%\\
Heterogeneous & $\delta=+0.1$ & 0.695 & 0.567 & 0.128 & +0.0013 [-0.0010,+0.0037] & 1.1\%\\
Heterogeneous & $\delta=+0.2$ & 0.693 & 0.567 & 0.126 & -0.0008 [-0.0045,+0.0028] & 2.0\%\\
Heterogeneous & Raw-DSIC & 0.992 & 0.476 & 0.516 & +0.2980 [+0.2754,+0.3195] & 26.6\%\\
\bottomrule
\end{tabular}
\caption{Complete decomposition for the prior-estimation-error magnitude grid. All rows use the same 1,000 paired cost draws per true profile as Table~\ref{tab:e2_misspecification_main}. Uniform denotes $k_i=1$ for all providers; heterogeneous denotes the provider-specific vector $k=(0.5,2,0.2,5,1)$ in Table~\ref{tab:e2_misspecification_main}. Only assumed shapes vary, with $\delta=\pm0.05,\pm0.1,\pm0.2$. Payment is $P=C_{\rm exec}+R$ before rounding. Payment changes and assignment disagreement are relative to correct-prior \textsc{Star}; intervals use 2,000 paired bootstrap resamples. Disagreement is the fraction of draws with at least one task assigned differently.}
\label{tab:e2_prior_error_grid_appendix}
\end{table*}

%% file: tables/e1_scaling_main.tex
\begin{table*}[t]
\centering
\normalsize
\begin{tabular}{rcccc}
\toprule
$|\mathcal V_T|$ & Completed instances & \shortstack{Task-assignment\\time (seconds)} & \shortstack{All-provider payment\\time (seconds)} & \shortstack{Maximum total\\time (seconds)}\\
\midrule
8 & 12/12 & 0.069 & 0.118 & 0.285\\
32 & 12/12 & 0.094 & 0.219 & 0.512\\
64 & 12/12 & 0.134 & 0.361 & 0.767\\
\bottomrule
\end{tabular}
\caption{Full-mechanism CPU scaling. Times aggregate the chain and original sparse-DAG generators, provider counts $N\in\{3,5\}$, and three seeds. Allocation and all-provider payment columns are medians; max total retains the slowest completed instance.}
\label{tab:e1_scaling_main}
\end{table*}

%% file: tables/e1_scaling_appendix.tex
\begin{table*}[t]
\centering
\normalsize
\begin{tabular}{lrrrrrr}
\toprule
family & $|\mathcal V_T|$ & $N$ & solved & allocation (s) & payments (s) & peak RSS (MiB)\\
\midrule
chain & 8 & 3 & 3/3 & 0.124 & 0.120 & 183.3\\
chain & 8 & 5 & 3/3 & 0.067 & 0.115 & 182.0\\
chain & 32 & 3 & 3/3 & 0.078 & 0.214 & 183.7\\
chain & 32 & 5 & 3/3 & 0.104 & 0.262 & 187.2\\
chain & 64 & 3 & 3/3 & 0.113 & 0.278 & 187.8\\
chain & 64 & 5 & 3/3 & 0.194 & 0.421 & 268.9\\
sparse DAG & 8 & 3 & 3/3 & 0.068 & 0.101 & 182.6\\
sparse DAG & 8 & 5 & 3/3 & 0.070 & 0.119 & 181.6\\
sparse DAG & 32 & 3 & 3/3 & 0.090 & 0.175 & 187.0\\
sparse DAG & 32 & 5 & 3/3 & 0.122 & 0.245 & 189.1\\
sparse DAG & 64 & 3 & 3/3 & 0.134 & 0.296 & 193.1\\
sparse DAG & 64 & 5 & 3/3 & 0.201 & 0.513 & 204.3\\
\bottomrule
\end{tabular}
\caption{Optimization runtime by graph family, task count, and provider count. Each row contains three fixed seeds. $N$ is the number of available providers; task eligibility follows the capability requirements. Memory is the maximum sampled sum of resident set sizes (RSS) of the Python worker and all solver processes it launches, sampled at 20-millisecond intervals.}
\label{tab:e1_scaling_appendix}
\end{table*}

%% file: references.bib
@inproceedings{hong2023metagpt,
  title={MetaGPT: Meta programming for a multi-agent collaborative framework},
  author={Hong, Sirui and Zhuge, Mingchen and Chen, Jonathan and Zheng, Xiawu and Cheng, Yuheng and Wang, Jinlin and Zhang, Ceyao and Wang, Zili and Yau, Steven Ka Shing and Lin, Zijuan and others},
  booktitle={The Twelfth International Conference on Learning Representations},
  year={2023}
}

@inproceedings{qian2024chatdev,
  title={Chatdev: Communicative agents for software development},
  author={Qian, Chen and Liu, Wei and Liu, Hongzhang and Chen, Nuo and Dang, Yufan and Li, Jiahao and Yang, Cheng and Chen, Weize and Su, Yusheng and Cong, Xin and others},
  booktitle={Proceedings of the 62nd Annual Meeting of the Association for Computational Linguistics (Volume 1: Long Papers)},
  pages={15174--15186},
  year={2024}
}

@article{
chen2024frugalgpt,
title={Frugal{GPT}: How to Use Large Language Models While Reducing Cost and Improving Performance},
author={Lingjiao Chen and Matei Zaharia and James Zou},
journal={Transactions on Machine Learning Research},
issn={2835-8856},
year={2024},
url={https://openreview.net/forum?id=cSimKw5p6R},
note={Featured Certification}
}

@article{ong2024routellm,
  title={Route{LLM}: Learning to route {LLM}s with preference data},
  author={Ong, Isaac and Almahairi, Amjad and Wu, Vincent and Chiang, Wei-Lin and Wu, Tianhao and Gonzalez, Joseph E and Kadous, M Waleed and Stoica, Ion},
  journal={arXiv preprint arXiv:2406.18665},
  year={2024}
}

@inproceedings{zhang2025let,
  title={Let the {LLM} Stick to Its Strengths: Learning to Route Economical {LLM}},
  author={Zhang, Yi-Kai and Lu, Shiyin and Chen, Qing-Guo and Luo, Weihua and Zhan, De-Chuan and Ye, Han-Jia},
  booktitle={The Thirty-ninth Annual Conference on Neural Information Processing Systems},
  year={2025}
}

@article{myerson1981optimal,
  title={Optimal auction design},
  author={Myerson, Roger B},
  journal={Mathematics of operations research},
  volume={6},
  number={1},
  pages={58--73},
  year={1981},
  publisher={INFORMS}
}

@inproceedings{wei2022chain,
  title={Chain-of-thought prompting elicits reasoning in large language models},
  author={Wei, Jason and Wang, Xuezhi and Schuurmans, Dale and Bosma, Maarten and Xia, Fei and Chi, Ed and Le, Quoc V and Zhou, Denny and others},
  booktitle={Advances in Neural Information Processing Systems (NeurIPS)},
  volume={35},
  pages={24824--24837},
  year={2022}
}

@article{yao2023tree,
  title={Tree of thoughts: Deliberate problem solving with large language models},
  author={Yao, Shunyu and Yu, Dian and Zhao, Jeffrey and Shafran, Izhak and Griffiths, Tom and Cao, Yuan and Narasimhan, Karthik},
  journal={Advances in neural information processing systems},
  volume={36},
  pages={11809--11822},
  year={2023}
}

@inproceedings{besta2024graph,
  title={Graph of thoughts: Solving elaborate problems with large language models},
  author={Besta, Maciej and Blach, Nils and Kubicek, Ales and Gerstenberger, Robert and Podstawski, Michal and Gianinazzi, Lukas and Gajda, Joanna and Lehmann, Tomasz and Niewiadomski, Hubert and Nyczyk, Piotr and others},
  booktitle={Proceedings of the AAAI conference on artificial intelligence},
  volume={38},
  number={16},
  pages={17682--17690},
  year={2024}
}

@inproceedings{wu2024autogen,
  title={Autogen: Enabling next-gen {LLM} applications via multi-agent conversations},
  author={Wu, Qingyun and Bansal, Gagan and Zhang, Jieyu and Wu, Yiran and Li, Beibin and Zhu, Erkang and Jiang, Li and Zhang, Xiaoyun and Zhang, Shaokun and Liu, Jiale and others},
  booktitle={First Conference on Language Modeling},
  year={2024}
}

@article{zaman2013combinatorial,
  title={A combinatorial auction-based mechanism for dynamic VM provisioning and allocation in clouds},
  author={Zaman, Sharrukh and Grosu, Daniel},
  journal={IEEE Transactions on Cloud Computing},
  volume={1},
  number={2},
  pages={129--141},
  year={2013},
  publisher={IEEE}
}

@inproceedings{singla2013truthful,
  title={Truthful incentives in crowdsourcing tasks using regret minimization mechanisms},
  author={Singla, Adish and Krause, Andreas},
  booktitle={Proceedings of the 22nd international conference on World Wide Web},
  pages={1167--1178},
  year={2013}
}

@book{nisan2007algorithmic,
  title={Algorithmic Game Theory},
  editor={Nisan, Noam and Roughgarden, Tim and Tardos, Eva and Vazirani, Vijay V}, 
  year={2007},
  publisher={Cambridge University Press}
}

@inproceedings{kwon2023efficient,
  title={Efficient memory management for large language model serving with pagedattention},
  author={Kwon, Woosuk and Li, Zhuohan and Zhuang, Siyuan and Sheng, Ying and Zheng, Lianmin and Yu, Cody Hao and Gonzalez, Joseph and Zhang, Hao and Stoica, Ion},
  booktitle={Proceedings of the 29th symposium on operating systems principles},
  pages={611--626},
  year={2023}
}

@article{zheng2023efficiently,
  title={Efficiently Programming Large Language Models using SGLang.},
  author={Zheng, Lianmin and Yin, Liangsheng and Xie, Zhiqiang and Huang, Jeff and Sun, Chuyue and Yu, Cody\_Hao and Cao, Shiyi and Kozyrakis, Christos and Stoica, Ion and Gonzalez, Joseph E and others},
  year={2023},
  publisher={arXiv}
}

@article{gim2024prompt,
  title={Prompt cache: Modular attention reuse for low-latency inference},
  author={Gim, In and Chen, Guojun and Lee, Seung-seob and Sarda, Nikhil and Khandelwal, Anurag and Zhong, Lin},
  journal={Proceedings of Machine Learning and Systems},
  volume={6},
  pages={325--338},
  year={2024}
}

@book{cygan2015parameterized,
  title={Parameterized algorithms},
  author={Cygan, Marek and Fomin, Fedor V and Kowalik, {\L}ukasz and Lokshtanov, Daniel and Marx, D{\'a}niel and Pilipczuk, Marcin and Pilipczuk, Micha{\l} and Saurabh, Saket},
  volume={5},
  number={4},
  year={2015},
  publisher={Springer}
}

@inproceedings{archer2001truthful,
  title={Truthful mechanisms for one-parameter agents},
  author={Archer, Aaron and Tardos, {\'E}va},
  booktitle={Proceedings 42nd IEEE Symposium on Foundations of Computer Science},
  pages={482--491},
  year={2001},
  organization={IEEE}
}

@inproceedings{yue2025masrouter,
  title={Masrouter: Learning to route llms for multi-agent systems},
  author={Yue, Yanwei and Zhang, Guibin and Liu, Boyang and Wan, Guancheng and Wang, Kun and Cheng, Dawei and Qi, Yiyan},
  booktitle={Proceedings of the 63rd Annual Meeting of the Association for Computational Linguistics (Volume 1: Long Papers)},
  pages={15549--15572},
  year={2025}
}

@inproceedings{yang2026bamas,
  title={BAMAS: Structuring budget-aware multi-agent systems},
  author={Yang, Liming and Luo, Junyu and Liu, Xuanzhe and Lou, Yiling and Chen, Zhenpeng},
  booktitle={Proceedings of the AAAI Conference on Artificial Intelligence},
  volume={40},
  number={35},
  pages={29802--29810},
  year={2026}
}

@inproceedings{chen2010frugal,
  title={Frugal mechanism design via spectral techniques},
  author={Chen, Ning and Elkind, Edith and Gravin, Nick and Petrov, Fedor},
  booktitle={2010 IEEE 51st Annual Symposium on Foundations of Computer Science},
  pages={755--764},
  year={2010},
  organization={IEEE}
}
